\documentclass{aa}

\usepackage{txfonts}
\usepackage{graphicx,amsmath,amssymb,multirow}
\usepackage[usenames,dvipsnames]{xcolor}

\usepackage{natbib}
\bibpunct{(}{)}{;}{a}{}{,}

\usepackage[colorlinks=true,citecolor=RoyalBlue,urlcolor=RoyalBlue,linkcolor=RedViolet]{hyperref}

\newcommand{\msun}{\ifmmode{{\rm M}_{\odot}}\else{${\rm M}_{\odot}$~}\fi}

\newcommand{\nable}{\boldsymbol{\nabla}}
\newcommand{\gcc}{{\rm g~cm}^{-3}}

\begin{document}

\title{
Protostellar birth with ambipolar and ohmic diffusion
}

\author{
N.~Vaytet\inst{1,2}, B.~Commer\c{c}on\inst{2}, J.~Masson\inst{3,2}, M.~Gonz\'{a}lez\inst{4}, G.~Chabrier\inst{2,3}
}

\authorrunning{N.~Vaytet et al.}
\titlerunning{Protostellar birth with ambipolar and ohmic diffusion}

\institute{
Centre for Star and Planet Formation, Niels Bohr Institute and Natural History Museum of Denmark, University of Copenhagen, {\O}ster Voldgade 5-7, DK-1350 Copenhagen K, Denmark\and
\'Ecole Normale Sup\'erieure de Lyon, CRAL, UMR CNRS 5574, Universit\'e de Lyon, 46 All\'ee d'Italie, 69364 Lyon Cedex 07, France \and
School of Physics, University of Exeter, Exeter EX4 4QL, UK \and
Univ Paris Diderot, Sorbonne Paris Cit\'e, AIM, UMR7158, CEA, CNRS, F-91191 Gif-sur-Yvette, France
\\ \email{neil.vaytet@nbi.ku.dk} 
}

\date{Received XX / Accepted XX}

\abstract
%Context
{The transport of angular momentum is fundamental during the formation of low-mass stars; too little removal and rotation ensures stellar densities are 
never reached, too much and the absence of rotation means no protoplanetary disks can form. Magnetic diffusion is seen as a pathway to resolving this 
long-standing problem.}
%Aims
{We aim to investigate the impact of including resistive magnetohydrodynamics (MHD) in simulations of the gravitational collapse of a 1~\msun gas 
sphere, from molecular cloud densities to the formation of the protostellar seed; the second Larson core.}
%Methods
{We used the adaptive mesh refinement code \texttt{RAMSES} to perform two 3D simulations of collapsing magnetised gas spheres, including self-gravity, 
radiative transfer in the form of flux-limited diffusion, and a non-ideal gas equation of state to describe $\text{H}_{2}$ dissociation which leads to 
the second collapse. The first run was carried out under the ideal MHD approximation, while ambipolar and ohmic diffusion was incorporated in the second 
calculation using resistivities computed from an equilibrium chemical network.}
%Results
{In the ideal MHD simulation, the magnetic field dominates the energy budget everywhere inside and around the first hydrostatic core, fueling 
interchange instabilities and driving a
low-velocity outflow above and below the equatorial plane of the system. High magnetic braking removes essentially all angular momentum from the second 
core.
On the other hand, ambipolar and ohmic diffusion create a barrier which prevents amplification of the magnetic field beyond 0.1 G in the first Larson 
core which is now fully thermally supported. A significant amount of rotation is preserved and a small Keplerian-like disk forms around the second core. 
The ambipolar and ohmic diffusions are effective at radii below 10 AU, indicating that a spatial resolution of at least $\sim$1 AU is necessary to investigate the 
angular momentum transfer and the formation of rotationally supported disks.
Finally, when studying the radiative efficiency of the first and second core accretion shocks, we found that it can vary by several orders of magnitude 
over the 3D surface of the cores.}
%Conclusions
{This proves that magnetic diffusion is a pre-requisite to star formation. Not only does it enable the formation of protoplanetary disks in which 
planets will eventually form, it also plays a determinant role in the formation of the protostar itself.}


\maketitle

%%%%%%%%%%%%%%%%%%%%%%%%%%%%%%%%%%%%%%%%%%%%%%%%%%%%%%%%%%%%%%%%%%%%%%%%%%%%%%%%%%%%%%%%%%%%%%%%%%%%%%%%%%%%%
%%%%%%%%%%%%%%%%%%%%%%%%%%%%%%%%%%%%%%%%%%%%%%%%%%%%%%%%%%%%%%%%%%%%%%%%%%%%%%%%%%%%%%%%%%%%%%%%%%%%%%%%%%%%%

\section{Introduction}\label{sec:intro}

Angular momentum transport, and its regulation through magnetic braking, is one of the most important, yet poorly understood, physical mechanisms in star 
formation \citep[e.g.][]{HennebelleCharbonnel2013}. Under the ideal magnetohydrodynamic (herafter MHD) approximation, magnetic fields typically observed in molecular clouds \citep{Crutcher2012} are 
powerful enough to remove all angular momentum from collapsing dense stellar progenitors; a problem known as the `magnetic braking catastrophe' 
\citep{Matsumoto2004,HennebelleFromang2008,HennebelleTeyssier2008,MellonLi2008,Commercon2010}. Angular momentum is needed to form protoplanetary disks around 
young stars, and three possible solutions are currently being investigated by theoretical studies to try and solve the magnetic braking puzzle.

The first invokes the omnipresent turbulence in the molecular clouds, which, through turbulent reconnection, is thought to effectively regulate the
concentration of magnetic flux and lead to the formation of protoplanetary disks
\citep{Santos-Lima2012,Santos-Lima2013,Leao2013,Lazarian2013,Joos2013}.
Indeed, the first numerical studies of low-mass star formation were carried out in a rather 
simplified set-up where the collapsing cloud was in solid body rotation, permeated by a uniform magnetic field. It has also been proposed that a disorganised 
field is simply less efficient at removing angular momentum from the system \citep{Seifried2013,Seifried2015}. The second solution is once again related to 
the simulation set-up; it is argued that the situation where the magnetic field direction is aligned with the parent body's rotation axis is a very special case, 
with its own peculiarities, and unlikely to happen in nature. While the alignment between magnetic field and large density structures in molecular clouds has 
been studied with recent observations \citep{Planck2016,Hull2017}, the spatial resolution does not allow to perform the same quantitative analysis at the cloud dense core 
level. It is however perfectly possible that rotation axis and magnetic field are misaligned, especially if the magnetization is weak \citep{Mocz2017,Hull2017}.  \citet{Hull2013} present dust-polarization observations towards 16 nearby low-mass protostars and conclude that their data are consistent with disks that are not aligned with the magnetic fields in the cores from which they formed. This scenario was investigated by several authors
\citep{HennebelleCiardi2009,Joos2012,Li2013,Krumholz2013,Masson2016} and was found to also be conducive to disk formation. Nevertheless,  we note that as the magnetic 
dissipation relies on numerical diffusion, these studies do not always yield resolution converged results in the ideal MHD framework.

Finally, resistive effects in the induction equation were suggested as a means to reduce the pile-up of magnetic field around the central object
\citep{Duffin2008,MellonLi2009,Krasnopolsky2010,Li2011,Machida2011,DappBasu2012}. The gas inside protostellar envelopes and protoplanetary disks is poorly 
ionised, and ion-neutral collisions, which act as a diffusive process in the MHD equations, are omnipresent. While in early 2D studies, neither 
ohmic nor ambipolar diffusion were able to circumvent the magnetic braking catastrophe without requiring abnormally large resistivities \citep{Krasnopolsky2010,Li2011}, more recent 3D
calculations have shown that magnetic diffusion with realistic resistivities can facilitate the formation of flat rotationally dominated structures, with 
radii of about 50-60 astronomical units (AU) \citep{Tomida2015,Tsukamoto2015a,Masson2016,Hennebelle2016}.\footnote{It is not clear why \citet{Krasnopolsky2010} and \citet{Li2011} were not able to form rotationally supported disks in their calculations. Possible reasons include that their models were only 2D, and did not incorporate self-gravity, although this has never been confirmed.} This third pathway provides a physical diffusion mechanism which 
does not depend on the numerical resolution or the orientation of the magnetic field, it is simply governed by the microphysics of molecular cloud.

The vast majority of the works listed above have studied the first hydrostatic core stage of star formation (scales of $\sim$10 AU), and very few have 
considered the scales typical of the protostellar seed; the second Larson core \citep[$<0.1$ AU;][]{Larson1969,Masunaga2000,Vaytet2013}. 
The first full 3D hydrodynamical simulations of the formation of the second Larson core were carried out by \cite{Bate1998}. Since then, only a limited number of 
studies have reached the second core stage, with different numerical methods (nested grid codes, smoothed particle hydrodynamics), incorporating increasingly 
complex microphysics including magnetic fields, radiative transfer, magnetic diffusion. We summarise the list of these papers in Table~\ref{tab:past_studies}.  
The recent works by \citet{Tomida2015}, using a nested-grid code, and \citet{Tsukamoto2015a}, using smoothed particle hydrodynamics, were the first ones to 
include radiative transfer coupled to MHD with both ambipolar and ohmic diffusion.\footnote{We note that \citet{Tomida2015} did not quite follow the evolution of the 
collapsing system all the way up to the formation of the second core.} Even more recently, \citet{Wurster2018} went a step further by adding the Hall effect in 
their calculations of the second core formation. To help establish theoretical results, it is crucial to verify computational results across different codes and 
numerical methods. This paper aims to do precisely this, expanding on the latest Japanese and British studies to strengthen the validity of the star formation 
process. We follow the gravitational collapse of a dense sphere of magnetised gas, from molecular cloud densities to the formation of the protostar, including 
ambipolar and ohmic diffusion. We compare the results to the classical ideal MHD (IMHD) framework, and illustrate why magnetic diffusion is of paramount 
importance in low-mass star formation.

\begin{table*}
\caption{3D numerical studies of the formation of the second Larson core.}
\centering
\begin{tabular}{lccccc@{~}c@{~}c}
\hline
\hline
\multirow{2}{*}{Reference} & Numerical   & Equation     & Radiative                          & Magnetic               & \multicolumn{3}{c}{Non-ideal MHD}               \\
                           & method      & of state     & transfer?                          & fields?                & Ohmic?             & Ambipolar?  & Hall?            \\
\hline
\citet{Bate1998}           & SPH         & barotropic   & \textcolor{red}{No}                & \textcolor{red}{No}    & \textcolor{red}{No}    & \textcolor{red}{No}    & \textcolor{red}{No}    \\
\citeauthor{Machida2006} (\citeyear{Machida2006,Machida2007,Machida2008,Machida2011})        & Nested grid & barotropic   & \textcolor{red}{No}                & \textcolor{green}{Yes} & \textcolor{green}{Yes} & \textcolor{red}{No}    & \textcolor{red}{No}    \\
\citet{Whitehouse2006}     & SPH         & H$_{2}$+H+He & \textcolor{green}{Yes} (FLD)       & \textcolor{red}{No}    & \textcolor{red}{No}    & \textcolor{red}{No}    & \textcolor{red}{No}    \\
\citeauthor{Saigo2008} (\citeyear{Saigo2006,Saigo2008})          & Nested grid & barotropic   & \textcolor{red}{No}                & \textcolor{red}{No}    & \textcolor{red}{No}    & \textcolor{red}{No}    & \textcolor{red}{No}    \\
\citet{Stamatellos2007}    & SPH         & H$_{2}$+H+He & \textcolor{green}{Yes} (cooling) & \textcolor{red}{No}    & \textcolor{red}{No}    & \textcolor{red}{No}    & \textcolor{red}{No}    \\
\citet{Bate2010,Bate2011}  & SPH         & H$_{2}$+H+He & \textcolor{green}{Yes} (FLD)       & \textcolor{red}{No}    & \textcolor{red}{No}    & \textcolor{red}{No}    & \textcolor{red}{No}    \\
\citet{Tomida2013}         & Nested grid & H$_{2}$+H+He & \textcolor{green}{Yes} (FLD)       & \textcolor{green}{Yes} & \textcolor{green}{Yes} & \textcolor{red}{No}    & \textcolor{red}{No}    \\
\citet{BateTricco2014}     & SPH         & H$_{2}$+H+He & \textcolor{green}{Yes} (FLD)       & \textcolor{green}{Yes} & \textcolor{red}{No}    & \textcolor{red}{No}    & \textcolor{red}{No}    \\
\citet{Tomida2015}         & Nested grid & H$_{2}$+H+He & \textcolor{green}{Yes} (FLD)       & \textcolor{green}{Yes} & \textcolor{green}{Yes} & \textcolor{green}{Yes} & \textcolor{red}{No}    \\
\citet{Tsukamoto2015a}      & SPH         & H$_{2}$+H+He & \textcolor{green}{Yes} (FLD)       & \textcolor{green}{Yes} & \textcolor{green}{Yes} & \textcolor{green}{Yes} & \textcolor{red}{No}    \\
\citet{Wurster2018}      & SPH         & H$_{2}$+H+He & \textcolor{green}{Yes} (FLD)       & \textcolor{green}{Yes} & \textcolor{green}{Yes} & \textcolor{green}{Yes} & \textcolor{green}{Yes}    \\
This work                  & AMR         & H$_{2}$+H+He & \textcolor{green}{Yes} (FLD)       & \textcolor{green}{Yes} & \textcolor{green}{Yes} & \textcolor{green}{Yes} & \textcolor{red}{No}    \\
\hline
\end{tabular}
\label{tab:past_studies}
\end{table*}

\section{Numerical method and initial conditions}\label{sec:method_and_init_cond}

\subsection{\texttt{RAMSES} with non-ideal MHD and flux-limited diffusion}\label{sec:method}

The simulations were carried out using a modified version of the adaptive mesh refinement (AMR) code \texttt{RAMSES} \citep{Teyssier2002,Fromang2006} which incorporates the effects of ambipolar and ohmic diffusion \citep{Masson2012}, and radiative transfer via a time-implicit flux-limited diffusion (FLD) approximation \citep{Commercon2011b,Commercon2014}. The governing equations are
\begin{gather}
\frac{\partial \rho}{\partial t} + \nable \cdot \left( \rho \mathbf{v} \right) = 0 \label{equ:mass_cons}\\[3ex]
\frac{\partial \rho \mathbf{v}}{\partial t} + \nable \cdot \left[ \rho \mathbf{v} \otimes \mathbf{v} + \left( p + \frac{|\mathbf{B}|^{2}}{2} \right) \mathbb{I} - \mathbf{B} \otimes \mathbf{B} \right] = - \rho \nable \Phi - \lambda \nable E_{\mathrm{r}} \label{equ:mom_cons}\\[3ex]
\frac{\partial E}{\partial t} + \nable \cdot \left[ \left(E + p + \frac{|\mathbf{B}|^{2}}{2}\right)\mathbf{v}  - \mathbf{B}(\mathbf{B} \cdot \mathbf{v}) \right. \nonumber\\
\left. +~\frac{\eta_{\mathrm{O}}c^{2}}{4\pi} (\nable \times \mathbf{B}) \times \mathbf{B} + \frac{\eta_{\mathrm{A}}c^{2}}{4\pi|\mathbf{B}|^{2}} [((\nable \times \mathbf{B})\times \mathbf{B})\times \mathbf{B}]\times \mathbf{B} \right] \nonumber\\
=  - \rho \mathbf{v} \cdot \nable \Phi - \lambda \mathbf{v} \cdot \nable E_{\mathrm{r}} - \kappa_{\mathrm{P}}\rho c \left( a_{\mathrm{r}} T^{4} - E_{\mathrm{r}}  \right) \label{equ:ener_cons}\\[3ex]
\frac{\partial \mathbf{B}}{\partial t} - \nable \times \left[ \mathbf{v} \times \mathbf{B} - \frac{\eta_{\mathrm{O}}c^{2}}{4\pi} \nable \times \mathbf{B}\right. \nonumber\\
\hspace*{0.125\textwidth} \left. - \frac{\eta_{\mathrm{A}}c^{2}}{4\pi|\mathbf{B}|^{2}} [(\nable \times \mathbf{B})\times \mathbf{B}]\times \mathbf{B} \right] = 0 \label{equ:induction}\\[3ex]
\nable \cdot \mathbf{B} = 0 \label{equ:zerodivB}\\[3ex]
\nabla^{2} \Phi = 4\pi G \rho \label{equ:poisson}\\[3ex]
\frac{\partial E_{\mathrm{r}}}{\partial t} + \nable \cdot (\mathbf{v} E_{\mathrm{r}} ) + \mathbb{P}_{\mathrm{r}} : \nable \mathbf{v} \nonumber\\
\hspace*{0.125\textwidth}= \kappa_{\mathrm{P}}\rho c \left( a_{\mathrm{r}} T^{4} - E_{\mathrm{r}}  \right) + \nable \cdot \left( \frac{c\lambda}{\rho \kappa_{\mathrm{R}}} \nable E_{\mathrm{r}}\right) \label{equ:erad_cons}~.
\end{gather}
The quantities are (in order of appearance): the gas density $\rho$, time $t$, the gas velocity $\mathbf{v}$, the gas pressure $p$, the magnetic field $\mathbf{B}$, the 
identity matrix $\mathbb{I}$, the gravitational potential $\Phi$, the radiative flux limiter $\lambda$, the radiative energy $E_{\mathrm{r}}$. The total gas energy is 
defined as $E = \epsilon + \rho \mathbf{v}\cdot\mathbf{v}/2 + \mathbf{B}\cdot\mathbf{B}/2$ where $\epsilon$ is the internal gas energy. $\eta_{\mathrm{O}}$ and
$\eta_{\mathrm{A}}$ are the ohmic and ambipolar magnetic resistivities, $\kappa_{\mathrm{P}}$ is the Planck mean opacity, $c$ is the speed of light, $a_{\mathrm{r}}$ 
is the radiation constant, while $T$ represents the gas temperature, $G$ is the gravitational constant, $\mathbb{P}_{\mathrm{r}}$ is the radiation pressure, and $\kappa_{\mathrm{R}}$ is the Rosseland mean opacity.

Equations~(\ref{equ:mass_cons}), (\ref{equ:mom_cons}), and (\ref{equ:ener_cons}) describe the conservation of mass, momentum, and energy, respectively. 
Equation~(\ref{equ:induction}) is the induction equation, (\ref{equ:zerodivB}) is the divergence-free condition, (\ref{equ:poisson}) is the Poisson equation for self 
gravity, and (\ref{equ:erad_cons}) is the conservation of radiative energy density.
In this work, we used the HLL Riemann solver for the MHD, and the Minerbo flux limiter \citep{Minerbo1978} for the FLD which is defined as
\begin{equation}
\lambda =
\begin{cases}
~2/(3 + \sqrt{9+12 R^{2}})   & \text{if}~~~~ 0 \leq R \leq 3/2\\
~(1 + R + \sqrt{1+2 R})^{-1} & \text{if}~~~~ 3/2 < R \leq \infty
\end{cases}
\end{equation}
where $R = |\nable E_{\mathrm{r}}|/(\rho\kappa_{\mathrm{R}}E_{\mathrm{r}})$. The radiation pressure is given by $\mathbb{P}_{\mathrm{r}} = \mathbb{D}E_{\mathrm{r}}$, and the Eddington tensor is
\begin{equation}
\mathbb{D} = \frac{1-\chi}{2} \mathbb{I} + \frac{3\chi-1}{2} \mathbf{n}\otimes\mathbf{n} ~,
\end{equation}
with $\chi = \lambda + \lambda^{2}R^{2}$ and $\mathbf{n} = \nable E_{\mathrm{r}} / |\nable E_{\mathrm{r}}|$ \citep{Levermore1984}.
The code incorporates the gas equation of state of \citet{Saumon1995}, and its extension to low densities \citep[see][]{Vaytet2013}, for a mixture of Hydrogen (73\%) and 
Helium (27\%, in mass). %, with a Helium mass concentration of 0.27.
The interstellar dust and gas opacities were taken from \citet{Vaytet2013}.
These comprise the dust opacities of 
\citet{Semenov2003} (assuming a 1\% dust content, by mass) at low temperatures (below 1500 K), the molecular gas opacities of \citet{Ferguson2005} for temperatures 
between 1500-3200 K, and the atomic gas opacities from the OP project \citep{Badnell2005} above 3200 K.
To aid the convergence of the implicit radiative transfer solver, we artificially limited the optical depth per cell to a minimum 
value of $10^{-4}$. When the gas is optically thin, it is not crucially important for the heating and cooling mechanisms whether the opticaly depth is $10^{-8}$ or $10^{-4}$, but we observed that choosing the latter can typically cut the number of iterations in the conjugate gradient solver by a factor of 4 or more. We show a validation of this acceleration scheme in Appendix~\ref{app:min_optical_depth}.

The magnetic resistivities were computed from a reduced chemical network including neutral and charged species, as well as dust grains, using an earlier version of the 
\citet{Marchand2016} model. It is in fact identical to the fiducial model of \citet{Marchand2016} (with a cosmic ray ionisation rate of $10^{-17}~\text{s}^{-1}$) 
for densities below $10^{-8}~\gcc$, but features a smooth decay in both
$\eta_{\mathrm{A}}$ and $\eta_{\mathrm{O}}$ beyond this point, following \citet{Machida2007} who use this to represent the thermal ionization of alkali metals, instead 
of taking into account the effects of grain evaporation, thermal ionisation of potassium, sodium, and hydrogen, and grain thermionic emission.
Using this tool, a three-dimensional table of density, temperature, and magnetic field dependent resistivities was computed. During the simulations, the resistivities in 
each grid cell were interpolated on-the-fly according to the local state variables, greatly reducing computational cost but implying thermodynamical equilibrium. The 
resistivities severely limit the integration timestep, and a stable super-time stepping method for ambipolar diffusion on an AMR grid with level-by-level sub-cycling is still lacking from the literature. To speed up the calculations, the timestep was prevented from going below a fraction of the ideal MHD timestep. It is 
taken to be the minimum of the three timescales
\begin{gather}
\Delta t_{\mathrm{ID}} = 0.8 \frac{\Delta x}{\sum_{i=x,y,z} u_{i} + |\mathbf{v}_{i}|}\nonumber\\
\Delta t_{\mathrm{O}} = \text{max}\left(0.1 \frac{\Delta x^{2}}{\eta_{\mathrm{O}}},\xi\Delta t_{\mathrm{ID}} \right)\label{equ:MHDtimesteps}\\
\Delta t_{\mathrm{A}} = \text{max}\left(0.1 \frac{\Delta x^{2}}{\eta_{\mathrm{A}}},\xi\Delta t_{\mathrm{ID}} \right)\nonumber ~,
% \Delta t_{\mathrm{A}} = \text{max}\left(0.1 \frac{|\mathbf{B}|^{2} \Delta x^{2}}{\eta_{\mathrm{A}} v_{\mathrm{A}}^{2}\rho},0.1\Delta t_{\mathrm{MHD}} \right)\nonumber ~,
\end{gather}
where $\Delta x$ is the cell size, $\xi = 0.1$, and 
\begin{equation}
u_{i} = \sqrt{ \frac{1}{2} \left(w^{2} + v_{\mathrm{A}}^{2}\right) + \frac{1}{2}\sqrt{ \left( w^{2} + v_{\mathrm{A}}^{2} \right)^{2} - 4w^{2}\frac{B_{i}^{2}}{4\pi\rho} } }
\end{equation}
is the fast magnetosonic speed in direction $i$, where $v_{\mathrm{A}} = \sqrt{|\mathbf{B}|^{2}/(4\pi\rho)}$ is the Alfv\'{e}n speed, and the sound speed
\begin{equation}
w = \sqrt{ \frac{\gamma p}{\rho} + \frac{4E_{\mathrm{r}}}{9\rho} }
\end{equation}
includes the contribution from the radiation pressure \citep[see][]{Commercon2011b}.
The idea is that the exact amount of magnetic diffusion included is not crucially 
important, as long as some diffusion is operating (see Appendix~\ref{app:dt_ministep} for more details). It is however necessary to compute the resistivity coefficients 
accurately with a chemical network, as in \citet{Marchand2016}, as the densities and temperatures at which they either rise or fall are important. The mesh refinement 
criterion was defined so that the local Jeans length was always sampled with a minimum of 32 cells everywhere in the computational domain. Initial tests with lower 
resolutions yielded spurious heating between the first and second core stages, due to inefficient cooling (see Appendix~\ref{app:resolution} and \citealt{Vaytet2016}).

\subsection{Simulation set-up}\label{sec:init_cond}

We adopt initial conditions similar to those in \citet{Commercon2010}. A magnetised isothermal sphere of molecular gas with quasi uniform density, rotating about the $z$-axis with 
solid body rotation, is placed in a surrounding medium a hundred times less dense with equal temperature. The sphere has a mass $M_{0} = 1~\msun$, a radius $R_{0} = 2753$ 
AU, and a temperature $T_{0} = 10$ K, for an initial ratio of thermal to gravitational energies of
\begin{equation}
\alpha=\frac{5 R_{0}k_{\mathrm{B}}T_{0}}{2GM_{0}\mu m_{\mathrm{H}}} = 0.28 ~,
\end{equation}
where $k_{\mathrm{B}}$ is Boltzmann's constant, $\mu$ is the mean molecular weight ($=2.31$ initially for the $\text{H}_{2}$+He mixture), and $m_{\mathrm{H}}$ is the hydrogen atomic mass.
The density in the domain is defined by
\begin{equation}
\rho =
\begin{cases}
~\rho_{0}\left[1 + \delta_{\rho} \cos \left(2\arctan\left(\frac{y}{x}\right)\right)\right] & \text{if}~~~~ r < R_{0}\\
~\rho_{0}/100 & \text{if}~~~~ r > R_{0}
\end{cases}
\end{equation}
where $\rho_{0} = 6.76 \times 10^{-18}~\gcc$ and includes an $m=2$ perturbation of amplitude $\delta_{\rho} = 0.1$, which has been used in many of our previous works to favour fragmentation in the collapsing system \citep[see][]{Commercon2008,Commercon2010}. The amount of rotation given to 
the cloud is parametrised according to the ratio of rotational to gravitational energies, which was chosen to be
\begin{equation}
\frac{R_{0}^{3}\Omega_{0}^{2}}{3GM_{0}} = 0.01 ~,
\end{equation}
where $\Omega_{0}$ is the angular velocity.
The strength of the magnetic field is defined in terms of the mass-to-flux ratio normalised by the critical value of stability for a
uniform sphere
\begin{equation}\label{equ:mu}
\mu = \frac{ \int_{0}^{R_{0}} dM \bigg/ \int_{0}^{R_{0}} d\phi_{\mathrm{B}}}{\left(M /\phi_{\mathrm{B}}\right)_{\mathrm{crit}}} = 4 ~,
\end{equation}
where $\phi_{\mathrm{B}} = \pi r_{\text{cyl}}^{2} B_{0}$ and $\left(M / \phi_{\mathrm{B}}\right)_{\mathrm{crit}} = \frac{0.53}{3 \pi}\left( \frac{5}{G} \right)^{1/2}$ 
\citep{Mouschovias1976} and $r_{\text{cyl}} = \sqrt{x^{2} + y^{2}}$ is the cylindrical radius.
The magnetic field is initially parallel to, and 
invariant along, the axis of rotation $z$. The field is stronger in a cylinder of radius $R_{0}$ (with the dense core at its centre) than in the surrounding 
medium, with $B_{z}(r_{\text{cyl}}<R_{0}) = B_{0} = 100^{2/3} B_{z}(r_{\text{cyl}}>R_{0})$, where the factor of 100 comes from the difference in density between the core and the surroundings
\citep[see][]{Masson2016}.
The base grid at the coarsest level counted $64^{3}$ cells, and an additional 21 AMR levels yielded a final effective resolution of $8\times 10^{-5}$ AU.

\section{Results}\label{sec:results}

We performed two simulations; the first using the ideal MHD approximation (\texttt{runID}), and the second including ambipolar and ohmic diffusion (\texttt{runAO}), requiring 40,000 and 180,000 CPU hours, respectively\footnote{The high cost for the non-ideal MHD simulation does not originate from a computationally expensive magnetic diffusion module, but comes primarily from a highly reduced integration timestep between the first and second collapse stages, as ambipolar and ohmic resistivities increase inside the first hydrostatic core (see equation~\ref{equ:MHDtimesteps}).}. In the remainder of this paper, we focus on describing the differences between the two models.

\subsection{Early evolution}\label{sec:early_evol}

\begin{figure}
\centering
\includegraphics[width=0.48\textwidth]{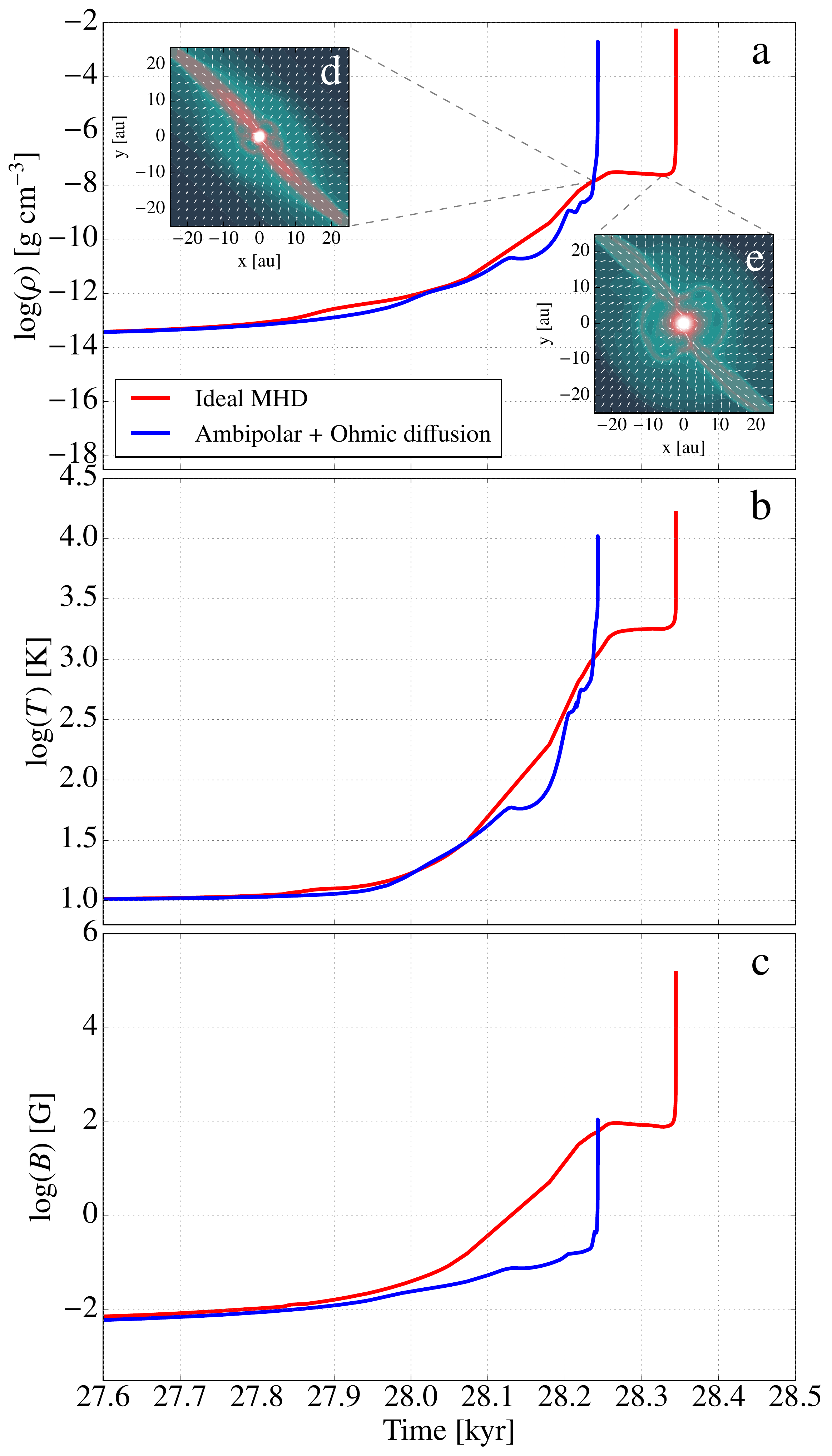}
\caption{Density (a), temperature (b) and magnetic field strength (c) as a function of time, for the densest cell in the system.
The red lines represent \texttt{runID}, while the blue lines are for \texttt{runAO}.
In the top panel, the two insets show maps of the logarithm of density in \texttt{runID} just before (d) and after (e) the development of the interchange instability (see text).}
\label{fig:early_evol}
\end{figure}

The evolution of a gravitationally collapsing dense molecular cloud core has been described in detail in past works 
\citep[see][for instance]{Masunaga2000,Vaytet2013}, and is displayed in Fig.~\ref{fig:early_evol} for our two runs.
It begins with an isothermal phase of contraction, clearly visible in the lower left corner of panel (b), where the compressive heating is lost via radiative cooling. 
As the density rises, the system's optical thickness increases and the radiative cooling becomes 
less and less efficient, until it can no longer counter-balance the compressive heating. The system enters its first adiabatic 
phase when densities exceed $\sim$10$^{-13}~\gcc$, where the first hydrostatic Larson core is formed.
The first core continues to accrete material from its envelope, and the 
sustained increase in mass forces the temperature to rise in the centre. When the gas reaches 2000 K, $\text{H}_{2}$ molecules begin 
to dissociate. The effective adiabatic index drops below the critical value of 4/3 for support against gravitational contraction, 
and a second, very rapid, phase of collapse takes place, at the end of which the second hydrostatic Larson core is formed.
The moment where the curves in all three panels exhibit a very sharp rise mark the onset of second collapse.

In the early stages ($t < 28230$ yr) \texttt{runID} and \texttt{runAO} have very similar central density and temperature evolutions. Only the strength of the 
magnetic field differs significantly already after 28000 yr, because the ambipolar and ohmic diffusion strongly hinders the condensation of
magnetic flux. Just before the second collapse in \texttt{runAO} ($t \simeq 28230$ yr), the discrepancy in $B$ has grown to almost 3 orders of magnitude. The effects of 
a strong field amplification are visible in the subsequent evolution of \texttt{runID}. All three displayed quantities show a plateau after
28250 yr, where contraction and heating is halted, delaying the second collapse. As illustrated by maps of the gas density in insets (d) and 
(e), this is caused by interchange instabilities that develop in the presence of extreme gradients in the magnetic field \citep{Spruit1995}. This effect was already observed in 
other works \citep[e.g.][]{Zhao2011,Tomida2015,Masson2016}, and is discussed further below.

\subsection{Physical picture at the time of second core formation}\label{sec:histograms}

We now turn to describing in more detail the properties of the first and second Larson cores, at a time right after the formation of the second core.
Finding a moment in both simulations where all aspects and structures of the collapsing systems can be directly compared is not trivial. The two runs reach the second core stage at slightly different times, and with different densities and temperatures in their centres. We defined the formation of the second core as the moment when a fully formed accretion shock is present, with a sharp density and velocity gradient at the core border. The justification for this somewhat arbitrary criterion will become clear in the following paragraphs. In addition,
in the remainder of this work,
a density threshold criterion -- favoured for its simplicity and robustness -- will be used to define the first and second Larson cores (see Appendix~\ref{app:cores}). All the cells with a density higher than
$10^{-10}~\gcc$ make up the first core, while the threshold is $10^{-5}~\gcc$ for the second core.

\begin{figure*}
\centering
\includegraphics[width=\textwidth]{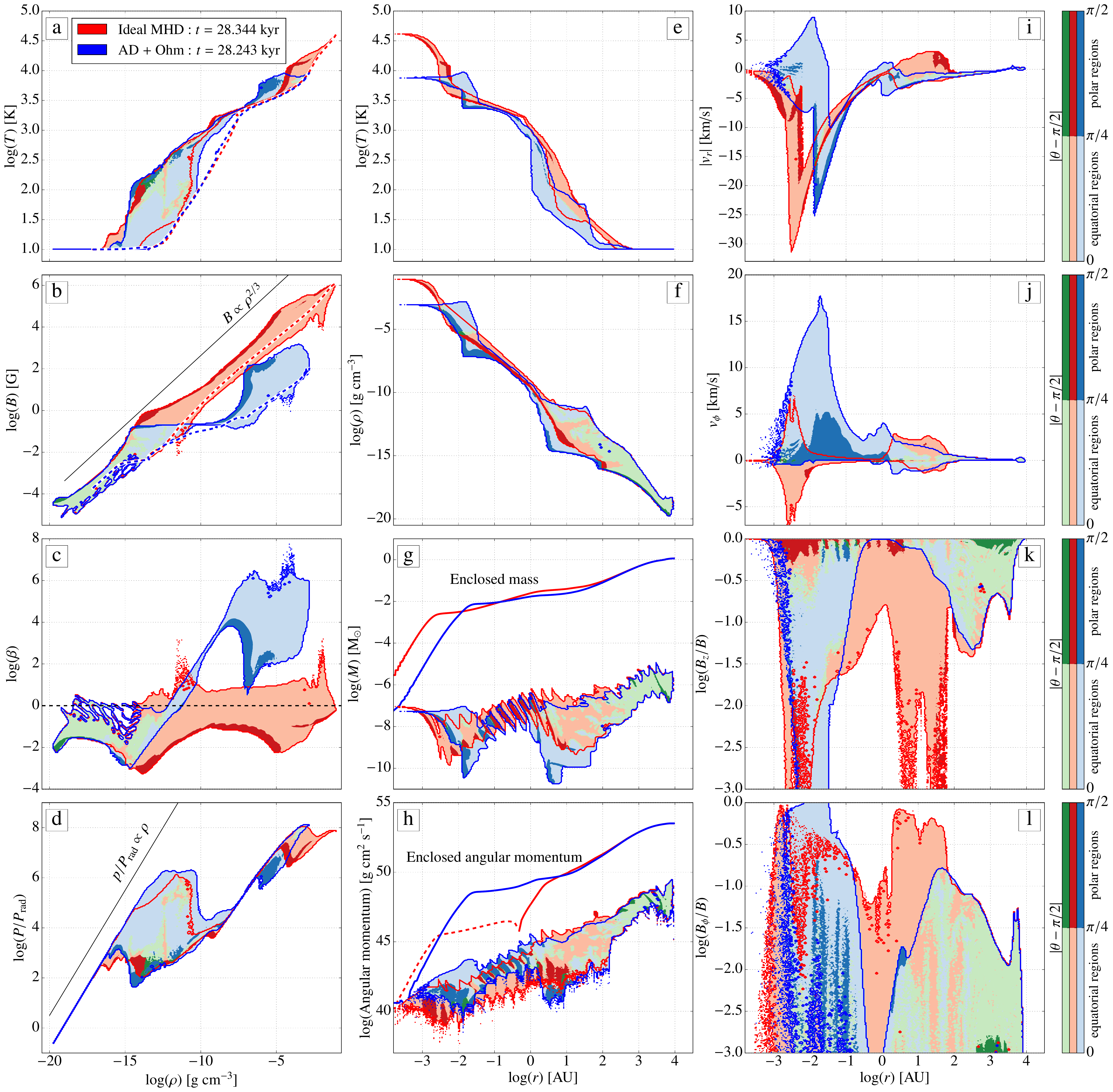}
\caption{\textbf{Left column}: Temperature (a), magnetic field (b), plasma $\beta$ (c), and ratio of thermal to radiative pressure (d) as a function of density, for every cell in the computational domain at the epoch of second core formation. The IMHD 
simulation is represented by the red colours, while the blue shades are for the NIMHD run. The green colours correspond to areas where IMHD and NIMHD results agree within 10\%. Each data set is delineated by a solid 
contour line which outlines the data distributions. The dark and light colours 
give an indication of the positions of the cells in the simulation box according to the $\theta = \cos^{-1}(z/r)$ angle: the light colours denote cells close 
to the equatorial region ($\pi/4 < \theta < 3\pi/4$) while dark colours show cells in the polar regions ($\theta<\pi/4$ or $\theta>3\pi/4$). The dashed lines in panels (a) and (b)
represent the time evolution of the central (densest) cell inside the mesh. The thin black line in panel (b) is the power law 
predicted from magnetic flux conservation in a contracting gas sphere. \textbf{Center and right columns}: radial distributions of various quantities for 
every cell in the computational domain. As in the left column, red colours are for \texttt{runID} while blue colours are for \texttt{runAO}. In panels (g) and (h) additional lines show the integrated enclosed mass and angular momentum, respectively, in successive spherical shells going outward from the centre of the system.}
\label{fig:histograms}
\end{figure*}

We first look at the evolution of the gas temperature at the centre of the system as a function of density, represented by the dashed lines in Fig.~\ref{fig:histograms}a.
The quasi isothermal contraction at low densities ($<10^{-13}~\gcc$) is clearly visible in the lower left corner.
The curves then follow an isentrope with an almost constant adiabatic index $\gamma_{\text{eff}} \simeq 7/5$\footnote{It is actually closer to 5/3 for 
$10^{-13} < \rho < 10^{-12}~\gcc$ \citep[see][]{Vaytet2014}.} until temperatures reach 2000 K and $\gamma_{\text{eff}}$ falls to $\sim$1.1, initiating the 
second collapse. The value of 7/5 is recovered towards the end of the tracks, once temperatures exceed $\sim$10$^{4}$ K.
The evolutions in \texttt{runID} and \texttt{runAO} are very similar, following tracks which strongly resemble the results of past 1-3D studies
\citep[][to only name a few]{Masunaga2000,Vaytet2013,Tomida2013,BateTricco2014}. The colour maps in Fig.~\ref{fig:histograms}a show a single 
snapshot in time of the distributions in the $(\rho,T)$ plane of all the cells in the simulation domain, just after the formation of the second Larson core. 
Red colours are for \texttt{runID} while blue is for \texttt{runAO}. The cells have been divided into two regions; the equatorial region (light colours) where the polar coordinate $\theta=\cos^{-1}(z/r)$ is in the range $\pi/4<\theta<3\pi/4$, and the polar region above and below the central protostellar object where $\theta<\pi/4$ or $\theta>3\pi/4$. The centre of the polar coordinate system is the centre of the second Larson core, found by calculating the mean coordinate of all cells with $\rho>10^{-5}~\gcc$.
The results from the two different calculations are overall qualitatively similar. The most noticeable difference is the density at which the shock heating occurs when the gas enters the second core. The shock heating happens at densities two orders of magnitude higher in \texttt{runID} than in \texttt{runAO}, suggesting that the protostellar core is more compact in the IMHD run. We also note that the gas in the polar regions (darker colours) undergoes shock heating earlier (i.e.~at lower densities) than around the equator (lighter colours), suggesting that the gas reaching the second Larson core is more diffuse close to the poles. This is actually visible below, in the density map around the second core in Fig.~\ref{fig:slices}r.

Figure~\ref{fig:histograms}b shows the distributions of the magnitude of the magnetic field vector $B = |\mathbf{B}|$ as a function 
of gas density. At low densities ($<10^{-15}~\gcc$), \texttt{runID} and \texttt{runAO} yield identical results. 
Above this point, we observe the same behaviour as in \citet{Masson2016}. While the magnetic field follows a $B\propto\rho^{2/3}$ 
power law in \texttt{runID} (consistent with magnetic flux conservation for a contracting gas sphere), a clear magnetic diffusion 
plateau appears in \texttt{runAO} around 0.1 G. This diffusion barrier strongly limits the amplification of the magnetic field, reduces 
magnetic braking, and prevents several IMHD peculiarities such as counter-rotation of gas inside the envelope surrounding the first 
core, or the development of interchange instabilities \citep[see][]{Masson2016}. As the resistivities begin to drop above densities 
of $\sim$10$^{-8}~\gcc$ (see Sec.~\ref{sec:method}), $B$ rises once again, but will remain between one and two orders of magnitude 
below the IMHD values. This has very important consequences for the properties of the second Larson core.

The ratio of thermal to magnetic pressure, otherwise known as the plasma $\beta = 2p/B^{2}$ is displayed in panel (c) as a function of density.
The effects of magnetic diffusion are once again unequivocal. At low densities, outside of the first core, the magnetic pressure dominates everywhere in both
\texttt{runID} and \texttt{runAO}. It also mostly dominates (or is comparable to the thermal pressure) inside the first and second cores in \texttt{runID}. 
However, the thermal pressure is orders of magnitude higher than the magnetic pressure when magnetic diffusion is included, as was reported in 
\citet{Masson2016}. The first and second hydrostatic cores are genuinely supported by thermal pressure, and the two simulations are forming two completely 
different protostars.

Panel (d) displays the ratio of thermal to isotropic radiative pressure $P_{\text{rad}}=E_{\text{r}}/3$, as a function of density. The two runs yield similar results. At low densities, radiative and thermal pressures are comparable, but as the gas contracts isothermally, $P_{\text{rad}}$ remains constant while $p$ scales linearly with density. As a result, the thermal pressure vastly dominates virtually everywhere in the collapsing system.

We now turn to studying in panels (e) to (l) the distributions of the fluid variables as a function of radius.
Panel (f) shows the gas density as a function of radius, and the distributions are relatively similar between IMHD and non-ideal MHD (NIMHD) models. The densities are in 
general lower along the polar directions than in the equatorial plane, which is expected for a disk forming in the plane of rotation. The 
second core in \texttt{runID} appears to be more compact than its \texttt{runAO} counterpart, and seems to also have a different structure; its density is 
relatively uniform, suggesting a more spherical morphology, while the \texttt{runAO} core is elongated in the equatorial plane and has density peaks away 
from the centre.
The temperature distribution in panel (e) shows again the more compact nature of the \texttt{runID} second core. It also reveals that in \texttt{runID}, 
temperatures are higher in most of the computational domain. This includes the regions inside the second core ($r<0.003$ AU), around the first core border 
($1<r<10$ AU) and also at larger radii ($r\sim100$ AU).

Panels (i) and (j) show the radial ($v_{r}$) and azimuthal ($v_{\phi}$) components of the gas velocity, as a function of radius. Two (negative) spikes in 
$v_{r}$ around 1 and 0.01 AU in \texttt{runAO} mark the first and second core borders, respectively. In \texttt{runID}, the first core border is less well defined and has a radius 3 times larger, while the second core is clearly visible around $3\times10^{-3}$ AU.  As expected, the highest velocities are found in the polar regions, 
where the gas is free-falling along the magnetic field lines, meeting no resistance along its path. The IMHD model has positive $v_{r}$ between 2 and 100 AU, 
representative of an outflow; a feature absent from \texttt{runAO}. The positive radial velocities inside the second core in \texttt{runAO} are a sign that 
the core is expanding because of strong rotation. Indeed, panel (j) shows a colossal amount of rotation in and around the \texttt{runAO} second core, while 
it is effectively zero in \texttt{runID}. The magnetic braking is so efficient in the latter that it has removed all angular momentum from the second core \citep[this confirms the results of][]{Tomida2013}.

Panels (k) and (l) display the vertical ($B_{z}$) and toroidal ($B_{\phi}$) components of the magnetic field, divided by the magnitude of the $B$ 
field vector. This reveals that around the first core region ($0.5<r<50$ AU), the field is much more vertical in \texttt{runAO} ($B_{\phi}$ falls to zero), 
while the opposite happens in \texttt{runID}. The magnetic diffusion allows the field lines to remain vertical without being drawn in by the fluid, unlike 
the IMHD model where perfect coupling between fluid and magnetic field means that the field lines are dragged into a pinched hourglass shape \citep[see][for 
example]{Krasnopolsky2010}, changing the orientation of the field and strongly reducing $B_{z}$.
The picture is almost reversed for the second Larson core, but for a different reason. The field is almost entirely toroidal in \texttt{runAO}
($B_{z}/B \rightarrow 0$ and $B_{\phi}/B \rightarrow 1$), because of the strong rotation of the gas which drags the field lines along (at these densities and 
temperatures, the gas is almost fully ionised and the field is once again perfectly coupled to the gas). On the other hand, $B_{\phi}$ remains rather small 
in \texttt{runID} because of the lack of rotation at the second core level. We also note that throughout the domain, the field remains mostly vertical in the 
polar regions, in both simulations, which is fully expected in a set-up where the rotation axis is initially aligned with the magnetic field.

Finally, in panels (g) and (h) we show the distribution of the mass and angular momentum, respectively, contained in the grid cells. The mass contained inside a cell may not provide much valuable information, as it is governed by our mesh refinement strategy, and the fact that it varies only lightly across the entire radial extent is simply a result of choosing to refine the grid according to the Jeans criterion. More interestingly, if we integrate the mass inside successive spherical shells around the protostar, we obtain the enclosed mass which we represent by the two solid lines in the upper half of panel (g). The two systems have similar mass profiles, apart from inside the second core which is more compact in \texttt{runID}. In the case of the angular momentum, the main difference between the two runs is a collection of cells in \texttt{runAO} with much higher angular momentum than in \texttt{runID}, in the range $-3 < \log(r) < -1.5$. This corresponds to the cells with high azimuthal velocities found in panel (j). As a consequence, the integrated angular momentum for radii below 1 AU is orders of magnitude higher in \texttt{runAO} than in \texttt{runID}. In fact, the exceedingly strong magnetic braking in \texttt{runID} even forced a sign reversal of the angular momentum inside the second Larson core (dashed red line). However, the amount of rotation is so small (see also Sect.~\ref{sec:morphologies}) that it is difficult to see as a bulk counter-rotating motion; the main component of the gas velocity is radially infalling at these radii.

This result is of crucial importance. It shows that magnetic diffusion (both ambipolar and ohmic) starts to become effective for radii below 10 AU, and 
even more so below 1 AU, indicating that a spatial resolution of at least $\sim$1 AU is necessary to correctly study angular momentum transfer and the formation of 
rotationally supported disks around protostars.\footnote{The maximum resolution of 0.15 AU in \citet{Masson2016} verifies this condition.}

\subsection{Morphologies}\label{sec:morphologies}

\begin{figure*}
\centering
\includegraphics[width=0.89\textwidth]{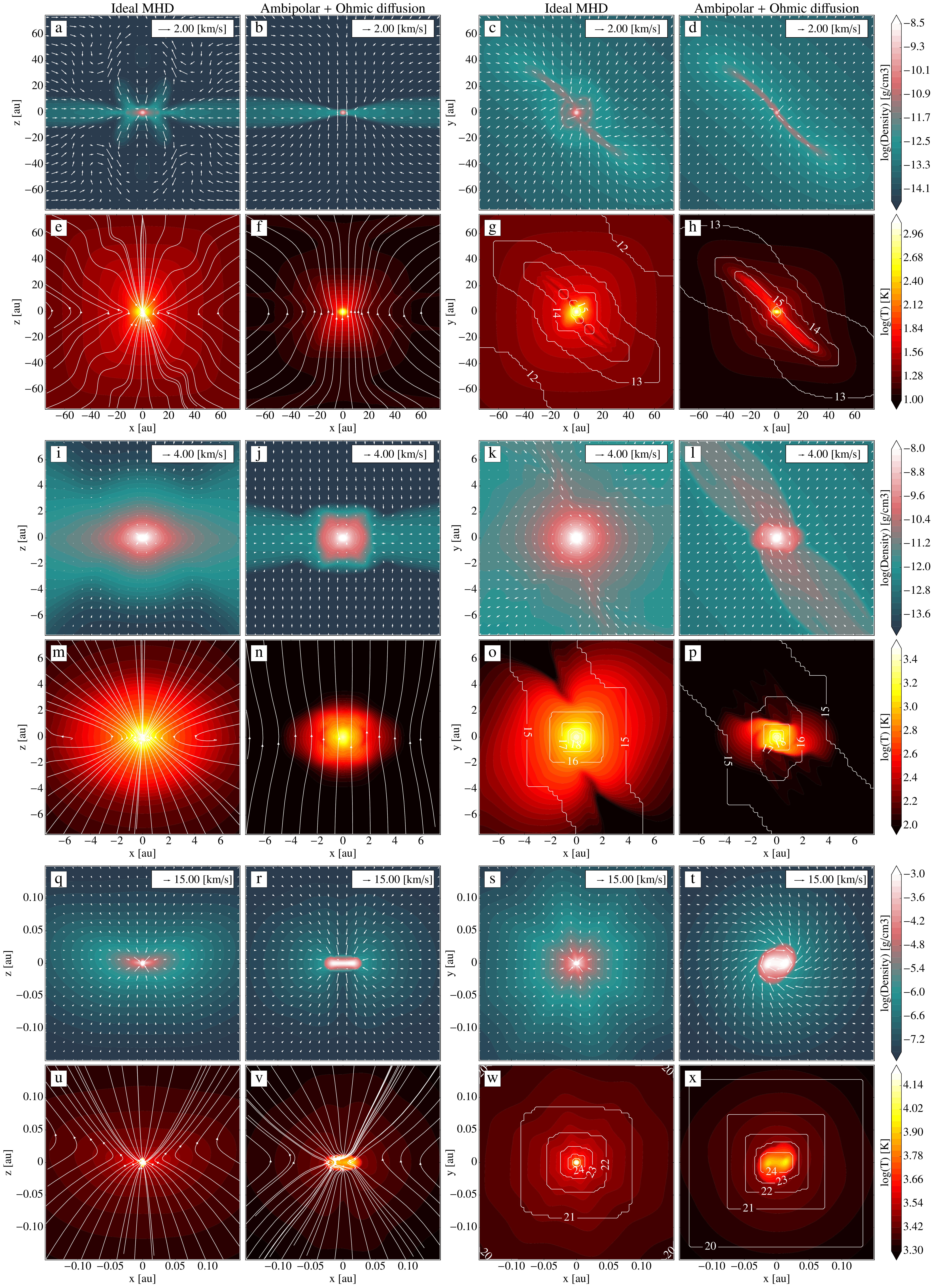}
\caption{Slices through the centre of the domain, comparing the morphologies of the protostellar system in \texttt{runID} (columns 1 \& 3) and \texttt{runAO} (columns 2 \& 4) on three different spatial scales at the epoch of second core formation. Panels (a)-(h) display a wide region around the first Larson core, the typical scale of a protoplanetary disk. Panels (i)-(p) show the immediate vicinity of the first Larson core. Panels (q)-(x) present the second Larson core and its close surroundings.
The two left columns show side $x$-$z$ views of the system, while the two right columns display the top $x$-$y$ perspective. The coloured maps in each row alternate between representing the gas density and temperature. The arrows on the density maps depict the gas velocity field. Overlayed onto the temperature maps are magnetic field lines (left column) and AMR level contours (right column).}
\label{fig:slices}
\end{figure*}

Figure~\ref{fig:slices} contains multiple slices through the data, comparing the morphologies of the protostellar system in \texttt{runID} (columns 1 and 3) and \texttt{runAO} (columns 2 and 4) on three different scales. The top two rows display a wide region around the first Larson core, the typical scale of a protoplanetary disk. The two middle rows show the immediate vicinity of the first Larson core, while the bottom two rows present the second Larson core and its close surroundings.
The two left columns show side $x$-$z$ views of the system, while the two right columns display the top $x$-$y$ perspective. The simulation times are the same as 
in Fig.~\ref{fig:histograms}.

\subsubsection{The first Larson core and its surroundings}\label{sec:morph_1stcore}

Panels (a)-(d) show gas density maps with velocity vectors. An equatorial density enhancement, typical of an accretion disk, is clearly 
visible in the side view of both simulations. In the top view, a filamentary structure extending from the north-west to the south-east of the 
protostar has formed from the initial density $m=2$ perturbation.\footnote{This may seem a little artificial but it in fact reproduces very well the density 
structures seen in simulations with more realistic turbulent initial conditions (Commercon et al.~2018, in~prep.).} A magnetic tower with outflowing velocity arrows (corresponding to the 
positive radial velocities in Fig.~\ref{fig:histograms}e) is observed in \texttt{runID} (a), while it is absent from \texttt{runAO} (b), as was the case in 
the strongly magnetised simulations of \citet{Masson2016}. Another large difference between the two runs, and another sign of strong amplification of the 
magnetic field, is the presence of `bubbles' in the $x$-$y$ view (c) of \texttt{runID} which are caused by interchange instabilities \citep[see][for a 
detailed study of these structures]{Zhao2011,Krasnopolsky2012}. While it has been argued that misalignement between the initial $B$ field and the rotation axis and 
turbulence are both able to prevent the formation of such structures \citep{Li2013,Li2014}, ambipolar and ohmic diffusion provide a physical rather 
than numerical diffusion that dominates the dissipation processes, with no dependence on the initial direction of the $B$ field nor the numerical resolution. 
The aligned case is no longer a special set-up with its strange behaviours and artefacts \citep[see][]{Masson2016}. Further evidence of the re-arrangement of
magnetic field lines provided by resistive effects is seen in the second row (panels e and f), where the magnetic field lines are very pinched 
in \texttt{runID}, while they are much more vertical in \texttt{runAO}. This corroborates our findings above; the field lines are no 
longer perfectly coupled to the gas and get less dragged in by the collapsing fluid. The modification of the magnetic field topology is provoked by the ambipolar
diffusion, the dominant mechanism in this region ($r<30$ AU; see Appendix~\ref{app:elsasser}).\footnote{This was once again already observed in the simulations of 
\citet[][see their Fig.~6]{Masson2016}.}
The temperature maps are also markedly different, with \texttt{runID} showing higher temperatures everywhere around the central protostar, up to a radius of 
$\sim$100 AU.

Taking a closer look at the first Larson core in panels (i) to (p), we notice that the disk is `puffed' up in \texttt{runID} (i) compared to \texttt{runAO}. The 
top view (k) also clearly show gas ejections from the interchange instabilities with outflowing velocity vectors. When looking at the time evolution of the gas 
temperature, we found that a sudden heating of the gas around the first core coincides with the development of the interchange instabilities, although we have 
not been able to establish if the instability is directly responsible for the heating. Other possible explanations include shock heating from waves launched by 
the instabilities, or irradiation from the protostar which is enhanced because the density -- and hence optical thickness -- of the gas around the first core 
drops as it gets ejected. One could even envisage a combination of the two, where shock heating raises the temperature around the core above $\sim$1000~K where 
dust grains start to sublimate, abruptly lowering the opacities, which in turn intensifies the irradiation.

In \texttt{runAO}, all the gas is moving towards the core, and the accretion is highly anisotropic, occuring primarily along the two high-density streams seeded 
by the perturbation in the initial conditions. In panels (m) and (n), the contrast in magnetic field orientation is glaring; the field in \texttt{runID} is 
pinched to the extreme, while it has become almost vertical in \texttt{runAO} due to the resistive effects. Panel (p) shows the high-density accretion streams 
hindering the propagation of heat from the central source, which progresses instead along the perpendicular direction. In \texttt{runID}, the more homogeneous 
density structure leads to a more homogeneous temperature distribution. The magnetic reconnection that occurs when interchange instabilities develop may also 
provide additional heating. However, this is not reconnection enabled by ohmic diffusion (that generates Joule heating) since it appears in the IMHD simulation; 
it is known as numerical reconnection. We have not been able to determine whether numerical reconnection heating is significant (or even happening at all) when 
compared to the irradiation from the central object, but the gas heating does appear to coincide with the development of the bubble-like ejections.

\subsubsection{The second Larson core}\label{sec:morph_2ndcore}

Panels (q) to (t) show once again density maps with velocity vectors for \texttt{runID} and \texttt{runAO}, but this time in the 
vicinity of the second Larson core. The morphologies are here also very different. The second core border is not very well defined in \texttt{runID}, where 
the gas density shows a rather smooth transition from $10^{-7}$ to $10^{-3}~\gcc$, as was already found in Fig.~\ref{fig:histograms}f. The protostellar seed also 
displays a loss of top-down symmetry (q), most probably due to magnetic flux re-distribution during the development of the interchange driven magnetic 
`bubbles'. We also note the absence of any rotation in panel (s), as already mentioned in Sec.~\ref{sec:histograms}. On the other hand, the \texttt{runAO} 
second core has a sharp border, strong rotation and a preserved top-down symmetry. It is flatter around the poles, due to both the rotation and the high 
infall speeds in the polar direction.
The top view (t) also reveals the early development of a spiral structure inside the core.
The second core masses for \texttt{runID} and \texttt{runAO} are $3.8\times10^{-3}~\msun$ and $7.4\times10^{-3}~\msun$, respectively.

\begin{figure}
\centering
\includegraphics[width=0.48\textwidth]{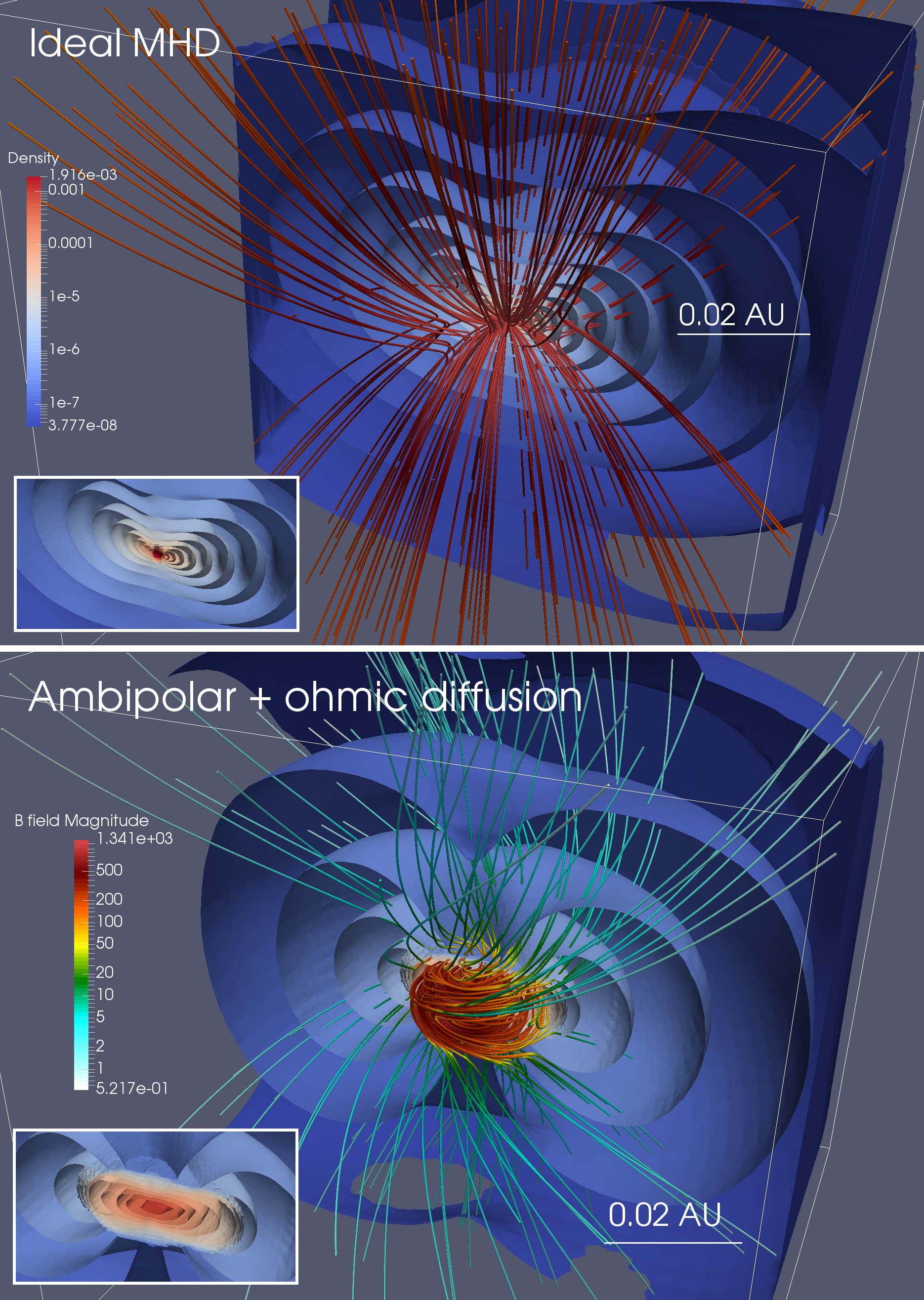}
\caption{3D visualization of logarithmically spaced density isosurfaces in the inner-most region of the computational domain showing the structure of the 
second Larson core, in the case of ideal (top) and non-ideal (bottom) MHD . The isosurfaces have been cut half-way in the $x$-direction. The magnetic field 
lines are overlayed, and have been coloured according to the magnitude of the magnetic field vector. The insets in the lower left corner of each panel show 
(with the same spatial scale) the central region of the system without the $B$ field for a better view of the morphology. The density and magnetic field 
colour scales apply to both panels.}
\label{fig:3d_view}
\end{figure}

The temperature maps with overlayed magnetic field lines in panels (u) and (v) expose the compact nature of the second core in \texttt{runID}. 
Temperatures at the very centre are higher than in \texttt{runAO}, and the core surroundings are also slightly warmer.
The field lines in the side view from 
both simulations have a very similar pinched shape, which is expected because the field is coupled to the gas in both runs as it is fully ionised at these 
scales.
It is always a challenge to view magnetic field lines in a 2D plane, and
Fig.~\ref{fig:3d_view} shows a 3D rendering of the magnetic field lines for both simulations, along with density isosurfaces. This view reveals the true 
topology of the field; a near perfect hourglass in \texttt{runID}, and strong winding inside the second core in \texttt{runAO}. The generation of toroidal 
field in \texttt{runAO} is expected to eventually lead to the launching of a fast outflow \citep{Machida2006,Tomida2013}.

\subsection{Late evolution}\label{sec:late_evol}

\begin{figure*}
\centering
\includegraphics[width=\textwidth]{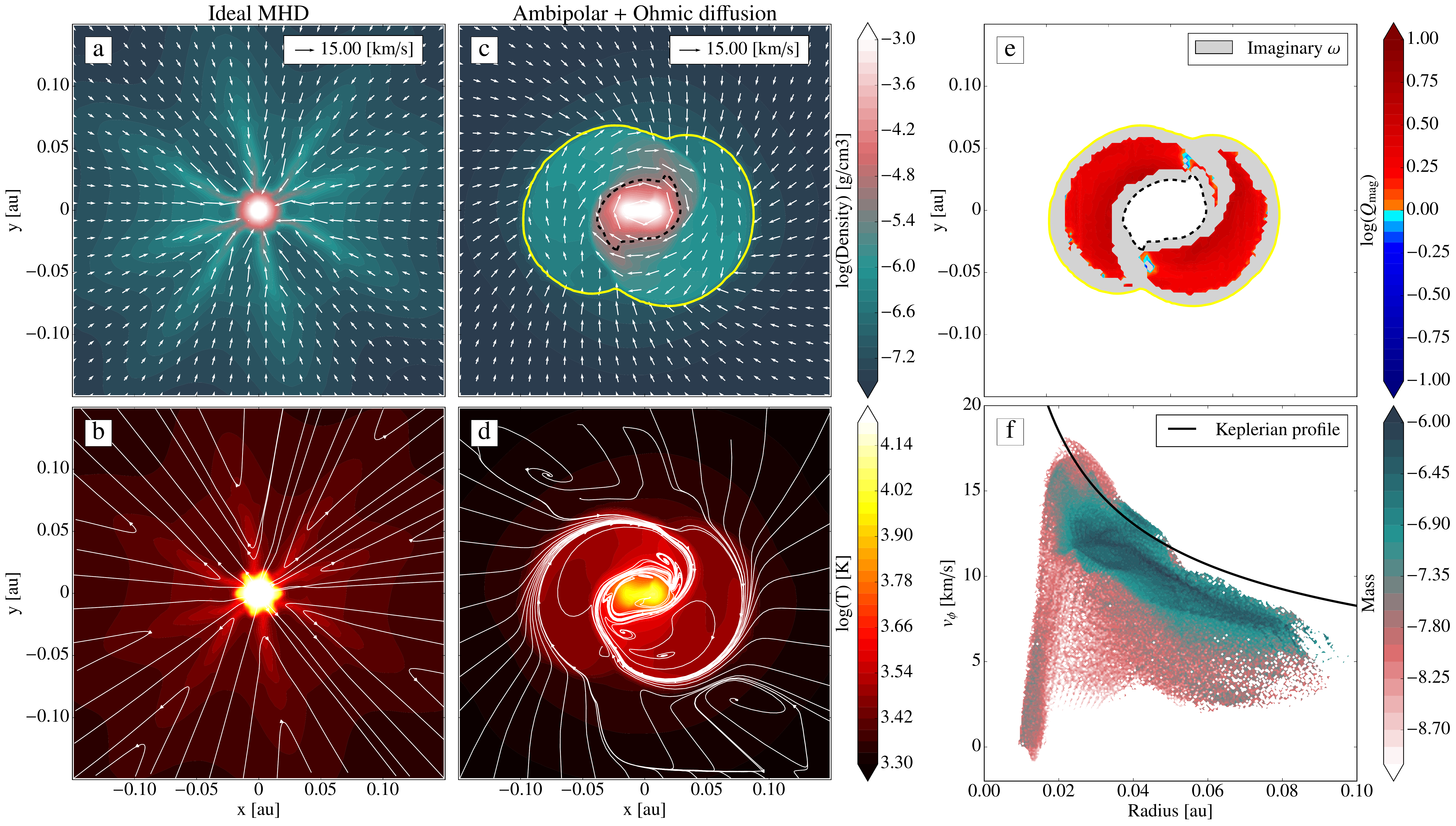}
\caption{Slices of the gas density with velocity vectors in \texttt{runID} (a) and \texttt{runAO} (c), about one month after the formation of the second Larson core. 
The area shown is the same as in Figs~\ref{fig:slices}s and \ref{fig:slices}t. In panel (c), the yellow contour marks the disk limit, taken as $\rho>10^{-6.7}~\gcc$, while the black dashed contour delineates the second hydrostatic core with $\rho>10^{-5}~\gcc$. Panels (b) and (d) show slices of the gas temperature with magnetic field streamlines overlayed. (e) Logarithmic map of the magnetic Toomre stability criterion $Q_{\text{mag}}$ inside the disk that forms around the second core in \texttt{runAO}. The grey-shaded areas indicate regions in the disk where the epicyclic frequency $\omega$ is imaginary and no $Q$ could be computed. The yellow and dashed black contours are the same as in panel (c). (f) Radial profile of the azimuthal velocity for all the cells inside the \texttt{runAO} second core disk. The colours code for the mass contained in a particular region of the plot. A Keplerian velocity profile is overlayed (black solid line).}
\label{fig:late_slice}
\end{figure*}

In this section, we look at the subsequent evolution of the IMHD and NIMHD systems.
Figure~\ref{fig:late_slice} shows density and temperature slices in the two simulations, approximately one month (24 days) after the formation of the second core. 
The second core in \texttt{runID} is still compact, has reached even higher densities and temperatures in its centre ($0.1~\gcc;10^{5}~\text{K}$), and 
appears to have filamentary accretion streams that are associated to the magnetic field topology (see panel b). Its mass is now $9.5\times10^{-3}~\msun$, with an 
effective mass accretion rate of $\sim$$7\times10^{-2}~\msun/\text{yr}$.

The small spiral instability in \texttt{runAO} detected in Fig.~\ref{fig:slices}t has developed into a small disk around the second core with two spiral 
arms. At this point, the second core mass has grown to $7.7\times 10^{-3}~\msun$, for an effective mass accretion rate of $\sim$$4\times10^{-3}~\msun/\text{yr}$ (the core is delineated by the black dashed contour in Fig.~\ref{fig:late_slice}c). It has a rotation period of $\sim$22 days. The disk mass is $1.8\times 10^{-4}~\msun$ (the disk was defined as the gas with densities in the range $10^{-6.7}~\gcc < \rho < 10^{-5}~\gcc$; this is marked by the yellow and dashed black contours).
We computed the magnetic Toomre stability criterion $Q_{\text{mag}}$ \citep{Kim2001} for this disk according to
\begin{equation}\label{equ:toomre}
Q_{\text{mag}} = \frac{\omega \sqrt{c_{\text{s}}^{2} + v_{\text{A}}^{2}}}{\pi G \Sigma} ~,
\end{equation}
where $c_{\text{s}}$ is the gas sound speed, $\Sigma$ is the disk surface density, and
\begin{equation}
\omega = \left( 4\Omega^{2} + 2\Omega r \frac{d\Omega}{dr} \right)^{1/2}
\end{equation}
is the epicyclic frequency of the gas with angular velocity $\Omega$. The surface density was integrated over the height of the disk, while $\omega$, $c_{\text{s}}$, and $v_{\text{A}}$ in equation~(\ref{equ:toomre}) actually represent the mass-weighted average values inside a given vertical
column (we note that $v_{\text{A}} \ll c_{\text{s}}$ because $\beta \gg 1$ at the densities considered). A map of $Q_{\text{mag}}$ is displayed in Fig.~\ref{fig:late_slice}e, revealing that the disk is stable against gravitational contraction. This is suggesting that forming tight binaries from fragmentation inside the second core disk may be difficult, but this is at such an early stage in the protostar's life that we cannot rule it out with the present result. Indeed, the disk is still rapidly growing in mass (see below), and may become unstable at a later stage.
In addition, Fig.~\ref{fig:late_slice}f shows the distribution of the azimuthal velocity as a function of radius of all the cells inside the disk. Even though the shape of the rotation profile is Keplerian-like, the disk is mostly sub-Keplerian, which is in agreement with the fact that the core is still accreting mass.
Finally, it should also be noted that our resolution is insufficient to correctly characterise the viscous dissipation inside the disk and adequately treat the protostellar core accretion shock cooling through the disk. Moreover, following the disk evolution for many orbital periods is computationally prohibitive (see below), and by limiting ourselves to such early epochs, we are not capturing the global disk cooling. These two mechanisms can affect the disk temperature and hence its dynamics and gravitational stability.

The very stringent Courant-Friedrichs-Lewy \citep[CFL;][]{Courant1967} condition inside the second core (because of the high sound speed) makes it very difficult 
to integrate for long periods of time after the second core formation. The simulation essentially `freezes' in time, as the timestep in a central 
region about 0.05 AU in diameter plunges to 10-20 s, which is not tracktable on astrophysical timescales.
In addition, the 27 levels of refinement needed to resolve the second core imply that the vast majority of cells lie in a tiny 
region in the centre of the simulation box, a situation where the CPU domain decomposition along a Hilbert space-filling curve performs poorly. Many 
processors end up holding no cells in the top AMR levels and spend much of their time waiting for the finer timesteps to complete on the other CPUs. 
Increasing the number of CPUs beyond 48 did not show convincing boosts in execution speeds, as any gain in processing power gets almost entirely counter-balanced 
by a heightened communications load.

\section{The first and second core accretion shocks}\label{sec:acc_shocks}

\begin{figure*}
\centering
\includegraphics[width=\textwidth]{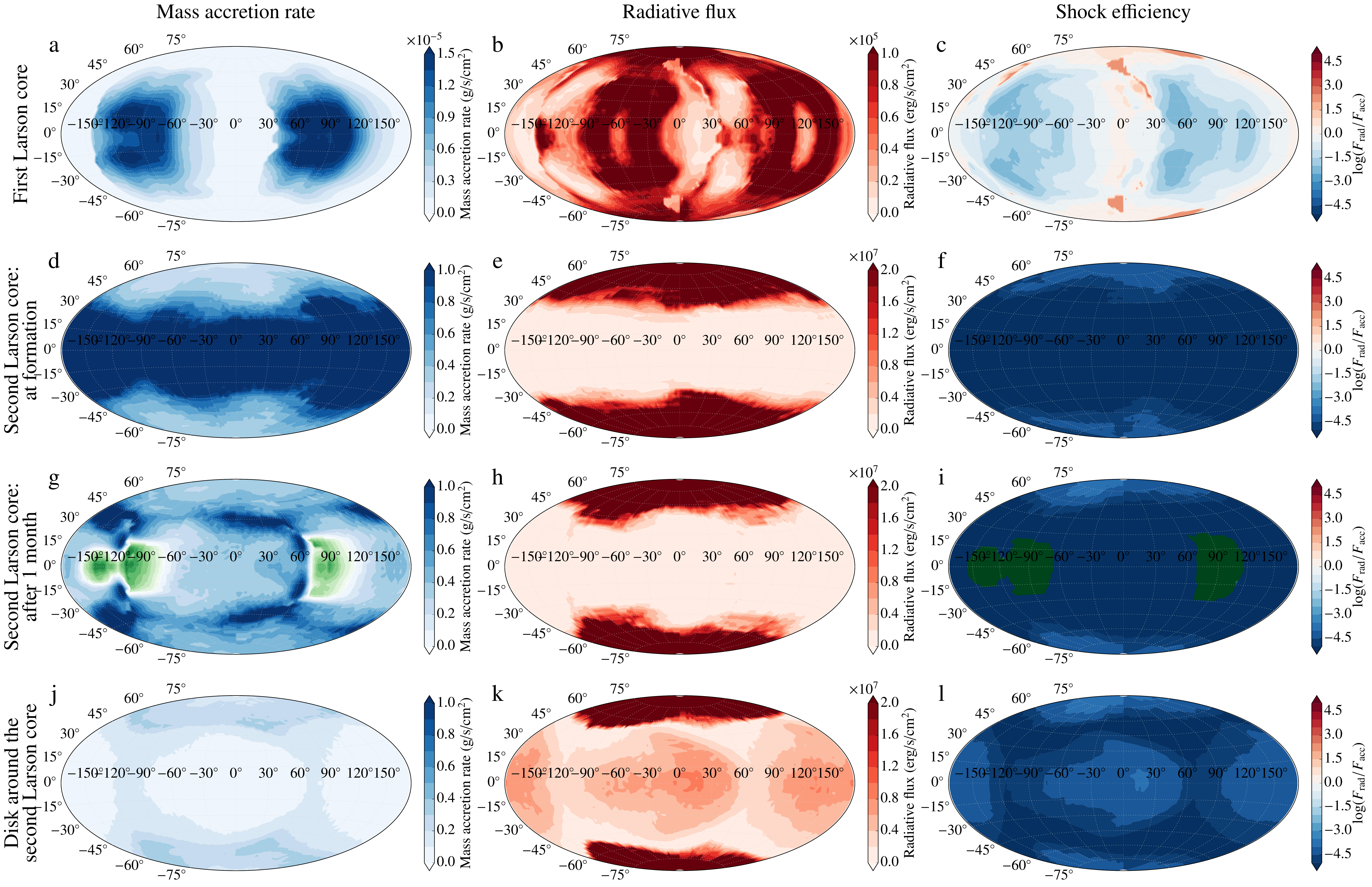}
\caption{Hammer projections (\texttt{runAO} only) of the mass accretion rate (left column), the radiative flux (middle column), and the ratio of radiative to 
accretion flux (right column). The first row is for the first Larson core, while the second row shows the second Larson core just after its formation. The third 
and fourth rows show the second Larson core 24 days after formation and the accretion flow at the edge of the disk around the second hydrostatic core, 
respectively. The green colours in panels (g) and (i) indicate negative values.}
\label{fig:shock_eff}
\end{figure*}

In this final section, we investigate in more detail the accretion flows onto the first and second Larson cores, and more particularly the radiative efficiency 
of the accretion shocks. Over the years, this subject has been of paramount importance to early evolutionary models of low-mass stars \citep[e.g.][]{Baraffe2012} 
as well as planets forming via the core accretion scenario \citep[e.g.][]{Mordasini2012}. Small changes in the fraction of the infalling gas energy that is 
either absorbed by the core, or radiated away at the accretion shock can yield significant differences in stellar and planetary luminosities and temperatures. 
However, the lack of accurate models of the accretion shocks which can predict the exact fraction of energy that is accreted or radiatied away in the literature 
have forced authors to bracket their results using two limiting cases known as `cold' (all energy is radiated away) and `hot' (all energy is absorbed) accretion. 
Recent numerical studies have suggested that the first Larson core accretion shock tends to be in the super-critical regime, radiating most of the infalling 
energy away \citep{Commercon2011b,Vaytet2012}, while the shock at the second Larson core border is sub-critical, transfering all the energy to the core 
\citep{Vaytet2013,Tomida2013}.

These predictions were mainly obtained with one-dimensional models of protostellar formation, and we now have the possibility to examine the 3D structure of the 
accretion flow and the resulting shock efficiency. Because it boasts the more complete microphysics, we consider only the \texttt{runAO} results in this section. In Fig.~\ref{fig:shock_eff}, panels (a), (b), and (c) show Hammer projections of the mass accretion rate per unit area
$\dot{M}$, the radiative flux, and the ratio of outgoing radiative flux to incoming gas energy flux $F_{\text{rad}}/F_{\text{acc}}$ just upstream of the first 
core accretion shock. Because the hydrostatic core is not spherical, we computed the maps by extracting density, velocity and radiative flux profiles along 
$64\times128$ different directions, starting from the centre of the second Larson core. The location of the accretion shock in each direction was chosen where 
the density and velocity gradients are at their maximum.
Equations~(\ref{equ:ener_cons}) and (\ref{equ:erad_cons}) give us the conservation of total, and radiative energy, respectively. Fig.\ref{fig:histograms}d 
revealed that at densities of the first and second Larson cores, the radiative energy is negligible compared to the gas internal energy, and we can thus drop the 
$\lambda \mathbf{v} \cdot \nable E_{\mathrm{r}}$ term in (\ref{equ:ener_cons}). In a similar manner, we drop all the terms involving the magnetic field because 
the plasma $\beta$ is above 100 for all densities above $10^{-10}~\gcc$ (see Fig.~\ref{fig:histograms}c).
In a purely conservative form, the gravity term in the right-hand side of equation~(\ref{equ:ener_cons}) should be included inside the left-hand side divergence. 
We re-write it as
\begin{equation}\label{equ:phi}
\rho \mathbf{v} \cdot \nable \Phi = \nable \cdot (\rho \mathbf{v} \Phi) - \Phi \nable \cdot (\rho \mathbf{v}) ~.
\end{equation}
Then, because we wish to look at a snapshot of the energy balance at the shock and not an evolution in time, we can assume a stationary state at the core 
accretion shocks, which means that (\ref{equ:phi}) reduces to $\nable \cdot (\rho \mathbf{v} \Phi)$ by virtue of (\ref{equ:mass_cons}), and can be inserted 
directly into the left-hand side divergence. We are now able to write the energy fluxes as
\begin{eqnarray}
F_{\text{acc}} & = & -v_{r} \left( \epsilon + \frac{\rho v_{r}^{2}}{2} + \frac{G M_{\text{enc}}}{r} \right) ~ds\\
F_{\text{rad}} & = & \frac{-c\lambda\nable E_{\mathrm{r}}}{\rho \kappa_{\mathrm{R}}}~ ds
\end{eqnarray}
where
$M_{\text{enc}}$ is the mass enclosed inside the sphere of radius $r$ and
$ds = r^{2}\sin\theta d\theta d\phi$ is the line of sight area element. Since we are computing an angular-dependent shock efficiency, we must measure it locally, 
rather than use a more global definition such as the energy balance scheme recently suggested by \citet{Marleau2017}.

Panel (a) reveals that mass accretion onto the first core is funneled along the dense filaments that we observed in Fig.~\ref{fig:slices}l. These appear 
as two large, almost circular, hot-spots in the panel (a) map, centred at longitudes of $75^{\circ}$ and $-105^{\circ}$. These regions dominate the 
total mass accretion rate, and illustrate once again that mass accretion is highly anisotropic. We also note that there are no negative values for
$\dot{M}$, meaning that radial velocities are negative everywhere; there are no outflows. In contrast, the radiative flux appears strong in regions of 
low accretion rate, although this is not a strict correlation. The radiation appears to propagate in directions where it meets low density gas which has 
a low optical depth. The resulting ratio of radiative to accretion flux in panel (c) is fascinating. Going against the commonly accepted paradigm that 
the core endures either cold or hot accretion, the map shows that it can be both at the same time. The accretion flux vastly dominates over its 
radiative counterpart (by 3 orders of magnitude) in the accretion hot-spots, while the two become comparable elsewhere. Going back to Fig.~\ref{fig:histograms}e,
we note that the temperature profile of \texttt{runAO} seems to show a temperature discontinuity for some of the gas at a radius of 1 AU, where the temperature
jumps from 100 to almost 1000 K. Such a discontinuity is indicative of a radiatively inefficient accretion shock, and the fact that this gas belongs to the 
equatorial regions (light blue colour) is consistent with the accretion hot-spots we report here. By contrast, there is a small dark blue (polar) region 
in the temperature profile of Fig.~\ref{fig:histograms}e around $\sim$3 AU that exhibits a less pronounced discontinuity, which corresponds to the radiatively 
efficient polar regions in Fig.~\ref{fig:shock_eff}c.

In the case of the second Larson core (panel d), the mass accretion rate is highest all around the equator, with no predominant hot-spots. This is a 
result of a higher gas density in the equatorial region just ahead of the shock (see Fig.~\ref{fig:slices}r). On the other hand, the radiative flux is 
higher in the polar regions where the lower density gas it has to travel through allows it to escape more freely. However, when we compare the accretion 
and radiative fluxes, even though we see structure dividing equatorial and polar regions, the accretion flux still dominates everywhere, by at least 4 
orders of magnitude. This result is thus in agreement with past 1-3D studies \citep{Vaytet2013,Tomida2013}. The surface integrated mass accretion rate 
is colossal at $0.28~\msun / \text{yr}$ \citep[also in agreement with][]{Vaytet2013}, and it is difficult to imagine that this will be sustained for very
long, as the protostar would finish 
accreting its entire 1~\msun envelope in under 4 years. Even though we have only run the simulation for $\sim$1 month after the formation of the second 
core, we already observe a dramatic drop in mass accretion rate in our final snapshot. Figure~\ref{fig:shock_eff}g displays the structure of the 
accretion flow onto the second core once the disk seen in Fig.~\ref{fig:late_slice} has formed; the strong equatorial accretion has disappeared and some 
regions of negative accretion (corresponding to positive values of $v_{r}$, shown in green) have even emerged. The disk acts as a buffer between the 
infalling material and the protostar; the gas is rotating in almost Keplerian fashion (see Fig.~\ref{fig:late_slice}) inside the disk, and radial inward 
motion is governed primarily by viscous transport. The radial velocity -- and hence the mass accretion rate -- at the protostellar surface is thus 
considerably reduced. For this final snapshot, we measure a surface integrated mass accretion rate of $0.074~\msun/\text{yr}$ onto the protostar, but 
neither this nor the initial mass accretion rate of $0.28~\msun / \text{yr}$ are a good indication of how fast the core is growing. Indeed, the 
accretion flow is unsteady and the average mass accretion rate during the first 24 days is only $4\times10^{-3}~\msun/\text{yr}$ (as mentioned in the 
previous section). Conversely, the mass accretion onto the disk is much more stable, with an average value of $2\times10^{-2}~\msun/\text{yr}$. It 
should however be noted that we probably do not have sufficient resolution to adequately resolve instabilities such as the magnetorotational instability 
\citep{Balbus1991}, which generate turbulence and regulate material and angular momentum transport inside the disk. Nevertheless, even if the mass 
accretion flow in unsteady, the ratio of infalling (kinetic and gravitational) to outgoing (radiative) energy is actually very stable; the accretion 
shock is radiatively inefficient throughout the early evolution of the protostar (panels f and i). The accretion energy flux also dominates over the 
radiation flux at the edge of the disk (panel l).

We emphasise here that these results only apply to the very early stages of the protostar's evolution, and cannot be assumed to hold for the remainder 
of the main accretion phase. They merely suggest that the second core accretion shock is initially radiatively inefficient, and reveal that it is 
possible to have both hot and cold accretion at the same time over the surface of the first core. We are reporting on the structure of the accretion 
flow at the birth of the protostar, and we do not know if this accretion arrangement can be applied to protostellar evolution models. We simply 
hint that the picture may not be either fully hot or cold; both regimes could be operating at the same time over the surface of the hydrostatic cores.

\section{Comparison with previous works}\label{sec:previous_works}

In this section, we compare the present study with previous articles that report on simulations of protostellar formation. For the sake of brevity, we limit 
ourselves to 3D non-ideal MHD simulations that have reached the second Larson core stage.

The first 3D models including ohmic diffusion were performed by \citet{Machida2006} using a nested-grid MHD code. The main difference between their models and 
our runs is that they use a barotropic equation of state, while we include radiative transfer via the FLD. They also lack ambipolar diffusion. Nevertheless, they 
already report a strong increase in plasma $\beta$ and angular momentum when number densities exceed $10^{14}~\text{cm}^{-3}$ in the resistive run compared to 
using ideal MHD. In the past five years, \citet{Tomida2013}, \citet{Tomida2015}, and \citet{Tsukamoto2015a} performed simulations including radiative transfer via the FLD, as well 
as non-ideal MHD with  ohmic diffusion and ambipolar diffusion. The most recent work by \cite{Wurster2018} includes radiative transfer and the three non-ideal MHD effects. 

Table \ref{tab:comparison} shows the properties of the first and second cores formed in our 
simulations. 
Overall, our results are qualitatively similar to those reported in the recent literature within a factor of a few (since we do not use the same definition 
criteria for the first and second cores,  we expect to have small differences). For instance, \citet{Tomida2013} reported second core mass of
$2\times10^{-2}~\msun$ one year after its formation. Assuming the system settles on timescales much shorter than a year after formation (i.e.~about a month, as 
observed in our simulation), this yields an average mass accretion rate of $2\times10^{-2}~\msun/\text{yr}$, which is five times our measured rate of
$4\times10^{-3}~\msun/\text{yr}$. However, \citet{Tomida2013} define their protostellar core as a pressure-supported body that would also include the small disk 
in our simulation (see Fig.~\ref{fig:cores}). Considering the disk as part of the second core means the second core mass accretion is now the flux at the disk 
border, which stands at $2\times10^{-2}~\msun/\text{yr}$ (cf.~Sec.~\ref{sec:acc_shocks}) and is now entirely consistent with \citet{Tomida2013}. The second core 
mass and size we derive are also roughly consistent with the results of \cite{Wurster2018} six months after the stellar core formation, who find masses of 
$1.5\times10^{-2}~\msun$ in IMHD and $3.4\times10^{-3}~\msun$ in NIMHD, as well as a radius of $0.013\times10^{-2}$~AU in both cases. We note that they use a similar 
criterion as ours for the second core definition, but with a density threshold a factor of ten higher. In addition, \citet{Tsukamoto2015a} found plasma beta within 
the first cores $\beta> 10^4$ in NIMHD and $\beta\sim 10$ in IMHD, which is fully consistant with Fig.~\ref{fig:histograms}c. 

Besides this qualitative agreement, there are some discrepancies in the structure of the collapsing core, as well as in the first core lifetime. First, 
\citet{Tomida2013} and \citet{Tomida2015} found that outflows and disks form early, even prior to the second collapse (with ohmic and ambipolar diffusion). The 
outflows reported in Tomida et al. have a relatively small extent 170~AU maximum at the end of the first core phase. Second they observed longer first core 
lifetimes and the latter increases when non-ideal MHD effects are included, whereas we find the opposite. In our models, we attribute this increase in the first 
core lifetime with IMHD to the development of interchange instabilities which heat up and bloat the first core (see Sec.~\ref{sec:early_evol}). Interchange 
instabilities are reported in \citet{Tomida2015} but do not affect the first core in IMHD as in ours. We think that these differences originate from the initial 
conditions. While we use uniform initial density profile, Tomida et al. used Bonnor-Ebert profile which is close to equilibrium. The time spent to form the first 
core is much longer when the initial core mass is close to the Bonnor-Ebert mass \citep[see][Fig.~7 therein]{Vaytet2016}. As previously mentioned, the accretion 
rate is a factor $\sim$5 higher in our models than in Tomida's, so that the first core evolves much quicker and the dynamic is more violent, leading to powerful 
magnetic interchange instability. The absence of outflows and large disks in our results is also consistent with the differences excepted between models using 
either a uniform or a Bonnor-Ebert density profile \citep{Machida2014}. In addition, \citet{Tsukamoto2015a} used uniform initial density and found that the 
protostellar disk forms after the second core in their NIMHD models. \cite{Wurster2018} also reports outflows at first core scales in NIMHD models using similar initial conditions as ours. However, they observe that outflows become broader and slower as the cosmic ray ionisation rate is reduced. The minimum ionisation rate they explore is $10^{-16}~\text{s}^{-1}$ while we use  $10^{-17}~\text{s}^{-1}$. Whether outflows launching at the first core scale depends on the cosmic ray ionsitation rate remains to be studied in detail. Clearly, the effect of the initial conditions, as well as the effect of the chemical set up used to estimate the MHD resistivities, has to be investigated in the near future to truly compare results.

\begin{table}
\caption{Properties of the first and second Larson cores extracted about one month after the birth of the second core.
}
{\centering
\begin{tabular}{@{}c@{~~~}c@{~~}c@{~~}c@{~~~}c@{~~~}c@{~~~}c@{~~}c@{~~}c@{~~~}c@{}}
\hline
\hline
\multirow{2}{*}{Model} & \multicolumn{3}{c}{$R_\mathrm{fc}$ (AU)} & $M_\mathrm{fc}$ & $\tau_\mathrm{fc}$ & \multicolumn{3}{c}{$R_\mathrm{sc}$ (AU)} & $M_\mathrm{sc}$  \\
                       & $x$  & $y$  & $z$                        &      ($\msun$)                                      &     (yr)                                      & $x$   & $y$  & $z$                       &      ($\msun$)                                      \\
\hline
\texttt{runID}         & 1.9 & 1.7 & 1.2	                  & 0.030                         & 239                                      & 0.023 & 0.020 & 0.021                      & 0.0095                         \\
\texttt{runAO}         & 1.5 & 1.1 & 1.1                       & 0.019                         & 129                                      & 0.028 & 0.028 & 0.012                       & 0.0077                         \\
\hline
\end{tabular}
\label{tab:comparison}}\\
{\footnotesize \textbf{Notes.} The columns are: first core radius (in the $x,y,z$ directions), mass, and lifetime, second core radius (in the $x,y,z$ directions), and mass.}
\end{table}

\section{Conclusions}\label{sec:conclusion}

We have performed two 3D simulations of the gravitational collapse of a dense sphere of molecular cloud gas. Both runs include the following physics: 
hydrodynamics, radiative transfer, self-gravity, a non-ideal gas equation of state, and magnetic fields. In the second run, the effects of ambipolar and 
ohmic diffusion were included in the MHD equations, and their impact on the simulation results were assessed through comparisons with the ideal MHD model.
The magnetic diffusion creates a barrier which prevents amplification of the magnetic field beyond 0.1 G in the first Larson core, with many consequences 
for the structure and evolution of the system. In the IMHD simulation, the magnetic field dominates the energy budget everywhere inside and around the first 
core, spawning interchange instabilities that create bubble-like ejections, as well as driving a low-velocity outflow above and below the equatorial plane of 
the system. A strong magnetic field also implies a heightened magnetic braking, removing essentially all angular momentum from the second Larson core.

When ambipolar and ohmic diffusion are present, the first and second cores become genuinely thermally supported and have a large amount of rotation. This leads 
to the formation of a small Keplerian-like gravitationally stable disk around the second core, and rolls the magnetic field lines into a toroidal topology which 
is expected to propel an outflow at the second core level. Due to stringent CFL limitations, it was however not possible for us to follow the evolution of the 
system long enough to observe the launch. We were also neither able to study the formation of a protoplanetary disk and a low-velocity outflow \citep{Gerin2017} 
around the first Larson core because the simulation essentially `froze' in time when the second core was formed. Future plans involve replacing the second core 
with a sink particle, allowing for much longer time integrations. The stark contrast between the ideal and NIMHD simulations proves that magnetic diffusion is of 
crucial importance to star-formation; not only does it enable the formation of disks in which planets will eventually form \citep{Masson2016}, it also shapes the 
protostar itself by preventing angular momentum loss and restoring thermal pressure support.

The use of idealised isolated initial conditions has been challenged by recent studies which claim that accretion processes in star formation are vastly 
influenced by the environment around the protostellar system \citep{Kuffmeier2017}. And while this may indeed be relevant at the first Larson core scale, we 
postulate that the dynamics at the second Larson core level are so disconnected, both in terms of spatial scales and evolutionary timescales, from the material 
100 AU away, that the impact of largescale turbulence would be negligible. Nevertheless, we are currently investigating the robustness of our results across 
different initial conditions, varying the parent cloud mass, changing the magnetic field strength and orientation, and introducing turbulence in the initial 
velocity field. Another shortcoming of the model presented in this paper is the lack of Hall effect in the MHD solver. Believed to be prominent in protoplanetary 
disks, the Hall effect has attracted much attention of late \citep[e.g.][]{Lesur2014,Tsukamoto2015b,Wurster2016,Tsukamoto2017}, and is considered to play a major 
role in angular momentum transport both inside the disk and in the protostellar envelope. We are in the process of implementing the Hall effect in our version of 
\texttt{RAMSES}. Last but not least, large uncertainties remain in the models used to estimate the resistivity coefficients because of poor constraints on the 
dust size properties (charge, size distribution) and on the chemistry at play in the high density and temperature regions of protostellar collapse. As a result,
it is currently not clear which non-ideal effects dominate in the different parts of the collapsing cloud, particularly for the Hall and ambipolar resistivities 
that strongly depend on the local physical and chemical conditions. Further work is required to better estimates of the non-ideal resitivities, which would in 
turn allow a more robust assessment of their impact on the star, disk, and planet formation process.

%%%%%%%%%%%%%%%%%%%%%%%%%%%%%%%%%%%%%%%%%%%%%%%%%%%%%%%%%%%%%%%%%%%%%%%%%%%%%%%%%%%%%%%%%%%%%%%%%%%%%%%%%%%%%
%%%%%%%%%%%%%%%%%%%%%%%%%%%%%%%%%%%%%%%%%%%%%%%%%%%%%%%%%%%%%%%%%%%%%%%%%%%%%%%%%%%%%%%%%%%%%%%%%%%%%%%%%%%%%

\begin{acknowledgements}
We are indebted to the anonymous referee for his/her insightful comments that have vastly improved the solidity of our study, with no stones left unturned.
We also thank Troels Haugb{\o}lle for very useful discussions during the writing of this paper.
NV gratefully acknowledges support from the European Commission through the Horizon 2020 Marie Sk{\l}odowska-Curie Actions Individual Fellowship 2014 programme (Grant Agreement no. 659706).
The research leading to these results has also received funding from the European Research Council under the European Community's Seventh Framework Programme (FP7/2007-2013 Grant Agreement no. 247060). We acknowledge financial support from "Programme National de Physique Stellaire" (PNPS) of CNRS/INSU, CEA and CNES, France. 
This work was granted access to the HPC resources of CINES (Occigen) under the allocation 2016-047247 made by GENCI.
We also made use of the astrophysics HPC facility at the University of Copenhagen, which is supported by a research grant (VKR023406) from Villum Fonden.
In addition, we thank the Service d'Astrophysique, IRFU, CEA Saclay, and the Laboratoire Astrophysique Instrumentation Mod\'{e}lisation, France, for granting us access to the supercomputer \texttt{IRFUCOAST} where the groundwork with many test calculations were performed.
All the figures were created using the \href{https://bitbucket.org/nvaytet/osiris}{\texttt{OSIRIS}}\footnote{\url{https://bitbucket.org/nvaytet/osiris}} visualization package for \texttt{RAMSES}, except Fig.~\ref{fig:3d_view} which was rendered with the \href{http://www.paraview.org/}{\texttt{PARAVIEW}}\footnote{\url{http://www.paraview.org}} software.
\end{acknowledgements}

%%%%%%%%%%%%%%%%%%%%%%%%%%%%%%%%%%%%%%%%%%%%%%%%%%%%%%%%%%%%%%%%%%%%%%%%%%%%%%%%%%%%%%%%%%%%%%%%%%%%%%%%%%%%%
%%%%%%%%%%%%%%%%%%%%%%%%%%%%%%%%%%%%%%%%%%%%%%%%%%%%%%%%%%%%%%%%%%%%%%%%%%%%%%%%%%%%%%%%%%%%%%%%%%%%%%%%%%%%%

\bibliographystyle{aa}

%%%%%%%%%%%%%%%%%%%%%%%%%%%%%%%%%%%%%%%%%%%%%%%%%%%%%%%%%%%%%%%%%%%%%%%%%%%%%%%%%%%%%%%%%%%%%%%%%%%%%%%%%%%%%
%%%%%%%%%%%%%%%%%%%%%%%%%%%%%%%%%%%%%%%%%%%%%%%%%%%%%%%%%%%%%%%%%%%%%%%%%%%%%%%%%%%%%%%%%%%%%%%%%%%%%%%%%%%%%

% \clearpage

\appendix

\section{Minimum optical depth per cell}\label{app:min_optical_depth}

\begin{figure}
\centering
\includegraphics[width=0.48\textwidth]{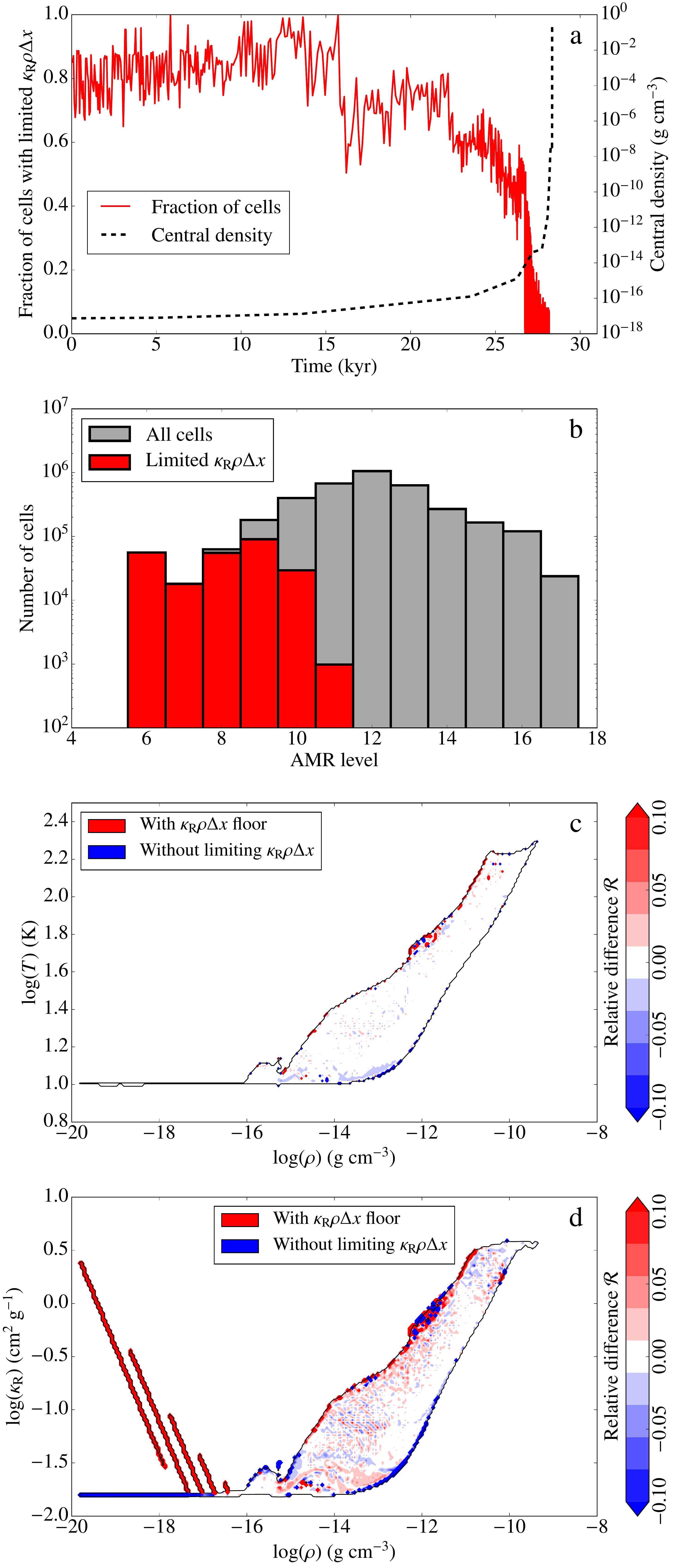}
\caption{(a) Fraction of cells inside the mesh where the optical depth is being limited as a function of time (red solid line). The dashed black line shows 
the density at the centre of the system as a function of time. (b) Number of cells in each level (grey) and the number of cells where the optical depth floor 
is operating (red), at a time of 28.180 kyr, when the first Larson core is formed. (c) Relative difference in 2D histograms of gas temperature as a function 
of density for all the cells in a simulation with optical depth limitation and a second simulation without, at $t = 28.180$ kyr. The colour scale gives a 
measure of $\mathcal{R} = N_{\text{limited}} / N_{\text{not limited}} - 1$, where $N_{\ldots}$ is the number of cells binned inside a $(\rho,T)$ pixel for 
the two different simulations. (d) Same as for (c) but in the case of the Rosseland mean opacity as a function of density.}
\label{fig:min_depth}
\end{figure}

In this section, we describe a scheme we devised to aid the convergence of the implicit radiative transfer solver. When the gas is optically thin, it is not 
crucially important for the heating and cooling mechanisms whether the optically depth inside a given cell is $10^{-8}$ or $10^{-4}$, as long as it is much 
less than unity. However, very low optical depths typically require many iterations for a time-implicit radiation solver to converge. We artificially limited 
the optical depth per cell to a minimum value of $10^{-4}$, by setting the mean Rosseland opacity to
\begin{equation}
\kappa_{\mathrm{R}} = \text{max}\left( \kappa_{\mathrm{R}}, \frac{10^{-4}}{\rho\Delta x}\right) ~.
\end{equation}
The flooring occurs in the large (low AMR level) low density cells, in the outer regions of the protostellar envelope. 
Figure~\ref{fig:min_depth}a shows the fraction of cells where the optical depth is being limited, with respect to the total number of cells in the 
simulation, as a function of time (red solid line). The black dashed line shows the evolution of the density at the centre of the collapsing cloud 
(i.e.~inside the densest cell) with time.
We see that while the fraction of cells with limited $\kappa_{\mathrm{R}}\rho\Delta x$ is large ($\sim$80\%) at early times, it drops below 0.1 when the 
first Larson core forms ($t \sim 28$ kyr and $\rho \sim 10^{-10}~\gcc$). In panel (b) of Fig.~\ref{fig:min_depth}, we show the total number of cells per AMR 
level (grey histogram), for a snapshot at a time of 28.180 kyr. The red histogram shows the cells where the optical depth is being limited. We can see that 
the floor is operating only in the outer layers of the collapsing system, from AMR level 6 to 11, and will not impact the properties of the first and second 
Larson cores.
In the ideal MHD simulation presented in the main part of this paper (up until a simulation time of 28.180 kyr), the total number of iterations is reduced by 
25\%, and the computational time reduced by 20\%.

To validate the optical depth flooring scheme, we show in panel (c) the temperature/density distribution of all the cells in the mesh for two simulations. 
The first has the optical depth limitation switched on, while it is turned off in the second. The coloured contours show the relative difference
$\mathcal{R}$ between the two simulations, for each $(\rho,T)$ pixel in the plot. It is defined as
$\mathcal{R} = N_{\text{limited}} / N_{\text{not limited}} - 1$, where $N_{\ldots}$ is the number of cells binned inside a $(\rho,T)$ pixel. A red area 
indicates that there are more cells from the simulation with the limitation scheme than from the run without the $\kappa_{\mathrm{R}}\rho\Delta x$ floor in 
that particular region of the plot, and vice-versa for blue areas.
The differences are expected to be the largest at low densities. However, in this isothermal phase of collapse, all the gas has a constant temperature of
10 K and the optical depth limiting scheme has basically no impact on the results. Small differences, of the order of 1\%, are visible at higher densities, 
but these mostly originate from the fact that the two simulation outputs have been written at slightly different times.\footnote{In \texttt{RAMSES}, outputs 
are only written when a coarse step has been completed, and it is often not trivial to write snapshots at exactly the same simulation time in two different 
simulations.}
Finally, in panel (d) we show the Rosseland mean opacity as a function of density, using the same convention as in panel (c). It is once again obvious that 
the limiter is only active in the outer layers of the infalling envelope, where the flow is still isothermal. The limited opacities show a stripy pattern 
which is due to the refinement of cells. We conclude that the optical depth limitation scheme does not appear to affect the thermodynamics of the system as 
it operates only in the isothermal stage of the collapse.

\section{The timestep limitation scheme}\label{app:dt_ministep}

One of the difficulties when working with diffusion processes on a mesh based framework is that the timestep criterion for numerical stability usually scales 
with the square of the mesh size $\Delta x$. This is indeed the case for ambipolar and ohmic diffusion, and is made worse by the fact that as densities 
increase, not only does $\Delta x$ decrease but the resistivities can also increase by several orders of magnitude \citep[see Fig.~5 in][]{Marchand2016}. 
This double effect (see equation~\ref{equ:MHDtimesteps}) causes the timestep $\Delta t$ to fall abruptly after the first Larson core is formed, and would 
require millions of timesteps to reach the second Larson core formation, making the problem non-tracktable. In the same spirit as limiting the optical depth 
per cell in the previous section, where we found that as long as the optical depth in a cell is much less than unity its exact value does not matter for our 
purposes, we postulate that as long as a strong magnetic diffusion is operating, the precise amount will not affect our results in a crucial way.

\begin{figure}
\centering
\includegraphics[width=0.48\textwidth]{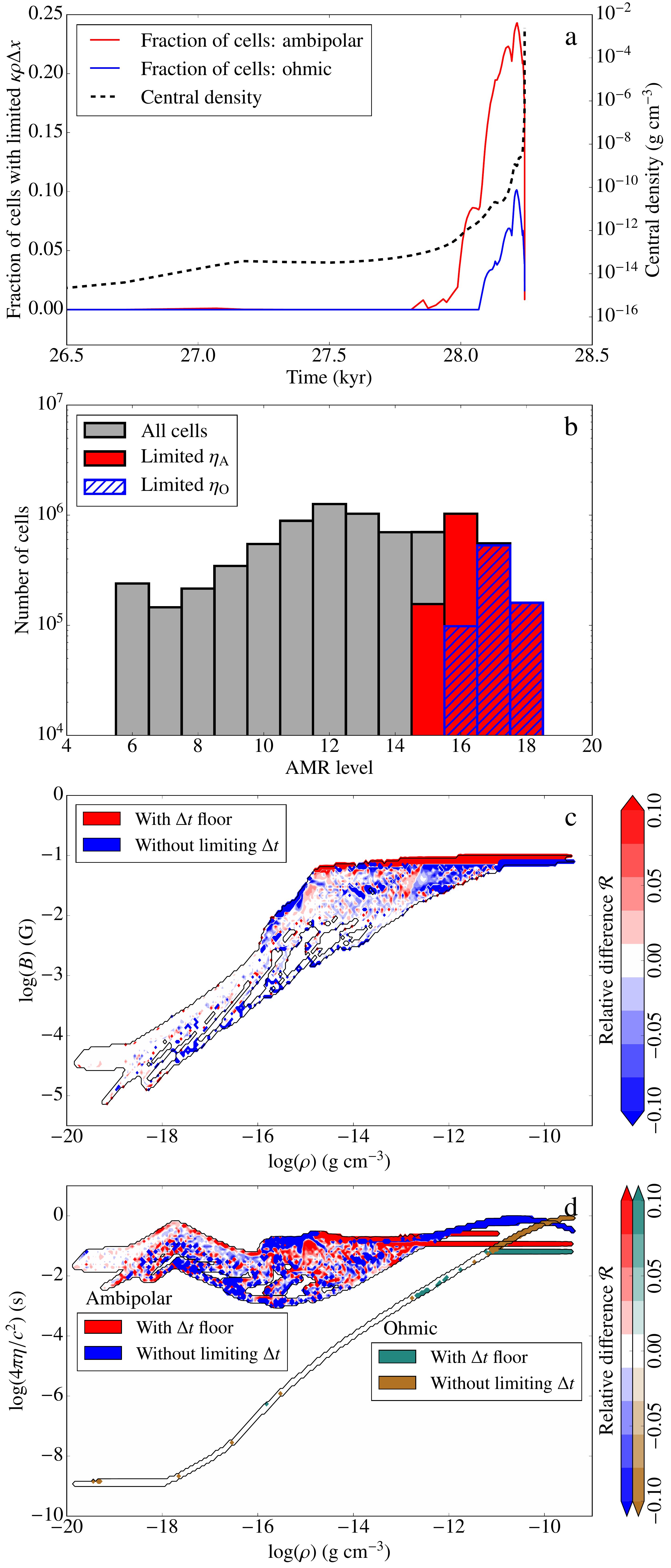}
\caption{(a) The fraction of cells inside the mesh where $\eta_{\text{A}}$ (red) and $\eta_{\text{O}}$ (blue) are being modified to prevent the MHD timestep 
from becoming too small, as a function of time. The dashed black line shows the evolution of the central density. (b) The number of cells per AMR level 
(grey) and the number of cells where the ambipolar (red) and ohmic (blue hatched) diffusion timestep floor is operating, at a time of 28.2 kyr. (c) Relative 
difference in 2D histograms of magnetic field strength as a function of density for all the cells in a simulation with $\Delta t$ flooring and a second 
simulation without, at $t = 28.2$ kyr. The colour scale is analogous to that of Fig.~\ref{fig:min_depth}. (d) Same as for (c) but in the case of the 
ambipolar (red-blue) and ohmic (green-brown) resistivities.}
\label{fig:brho_ministep}
\end{figure}

As mentioned in Sect.~\ref{sec:method_and_init_cond}, the method we have chosen to try and prevent the MHD timestep from reaching prohibitively low values is 
to artificially limit the value of $\Delta t$ to a fraction $\xi$ of the ideal MHD timestep $\Delta t_{\text{ID}}$. In practice, we found that setting the lower 
limit to $\xi=0.1$ was a good compromise between speedup and accuracy of results. We emphasise that we have no physical justification for the value of 0.1, it was simply chosen after months of testing. To ensure consistency between the imposed value of
$\Delta t$ and the magnetic diffusion, one has to artificially lower the resistivities in the cells which would have
$\Delta t_{\text{O,A}} < \xi \Delta t_{\text{ID}}$. The resistivities are thus overwritten with
$\eta_{\text{O,A}} = \text{min} \left( \eta_{\text{O,A}} , \frac{0.1 \Delta x^{2}}{\xi \Delta t_{\text{ID}}}\right)$. Note here that the factor of 0.1 in the numerator of the fraction on the right hand side is different from the $\xi =0.1$; it corresponds to the CFL-like factor that is used to compute the diffusion timestep, taken as a tenth of the time it would take for all the magnetic field inside the cell to diffuse.

Validation of this acceleration scheme is explicited in Fig.~\ref{fig:brho_ministep}. Panel (a) shows the fraction of cells inside the computational domain 
where the resistivities are being modified, as a function of time. The black dashed line represents the evolution of the central density, and we can see that 
as it reaches values characteristic of the first Larson core ($\sim10^{-12}-10^{-10}~\gcc$), the numbers of cells where $\Delta t_{\text{O,A}}$ is floored 
begin to increase. However, these fractions remain small throughout the simulation, peaking at 25\% for the ambipolar diffusion (red) and 10\% for the ohmic 
diffusion (blue). In addition, the flooring is only important during a transition phase between the formation of the first and second Larson cores, since 
after having increased with density, the resistivities begin to fall again once temperatures increase beyond $\sim$1500 K where the dust grains evaporate 
\citep[see Fig.~\ref{fig:histograms}b and][]{Marchand2016}. This is indeed reflected by the sharp fall in fractions (blue and red lines) as the density 
abruptly increases past $10^{-8}~\gcc$. A histogram showing the number of cells affected by the $\Delta t$ flooring for each AMR level, taken at a time of 
28.2 kr where the fractions in panel (a) reach their maxima, is displayed in panel (b). As the flooring operates only in the densest parts of the system, 
only the highest AMR levels are affected.

The resistivities affect primarily the magnetic field, and we show in panel (c) a distribution of the magnetic field as a function of density in every cell 
in two different simulations. The first has the acceleration scheme switched on, while the other is without. Because of the prohibitively small values of
$\Delta t$ in the simulation without timestep acceleration, we ran both calculation with a resolution of only 12 points per Jeans length. As in the previous 
section, the coloured contours show the relative difference $\mathcal{R}$ between the two simulations, for each $(\rho,B)$ pixel in the plot. It is defined 
as $\mathcal{R} = N_{\text{accel}} / N_{\text{no accel}} - 1$, where $N_{\ldots}$ is the number of cells binned inside a $(\rho,B)$ pixel. A red area 
indicates that there are more cells from the simulation with the acceleration scheme than from the run without the $\Delta t$ floor in that particular region 
of the plot, and vice-versa for blue areas. As expected, the timestep limitation scheme changes the magnetic diffusion plateau at high densities
($\rho > 10^{-13}~\gcc$), but only in a very minor way. The accelerated simulation still displays a strong magnetic diffusion barrier around 0.1 G, and the 
values of $B$ differ by 5\% or less in the rest of the computation box, compared to the run with the correct $\Delta t$.\footnote{Many of these errors are 
also due to the fact that the snapshots from the two simulations are not written at exactly the same simulation time.} This, we argue, is the justification 
for using the acceleration scheme; the magnetic diffusion is still operating, and still dominates over any numerical diffusion. The diffusion is crucial to 
limiting the magnetic braking and the accumulation of magnetic flux, and this is still achieved in the accelerated run. 
In the last panel (d), we show for 
informative purposes the values of the resistivities as a function of density, using the same colour convention as in panel (c). The differences below
$\rho \sim 10^{-13}~\gcc$ are once again due to a different simulation time output, and the resistivities are only modified by the acceleration scheme at high 
densities. Even though the resistivities can be modified by more than an order of magnitude, as long as they are high enough, the exact values of
$\eta_{\text{A,O}}$ do not seem to be important in the scope of our simulations.

It is of course difficult to predict the impact of such an acceleration scheme on simulation results without running the full (non $\Delta t$-limited) 
simulation first, as it is potentially highly problem-dependent.
Even though we tested the method across a range of initial conditions (different parent cloud masses, initial magnetization, temperature, rotation) and it 
always gave excellent results, we limited ourselves to the problem of a gravitationally collapsing magnetised body, and we must advise caution when using it 
for a different kind of set-up. 

\section{Resolution study}\label{app:resolution}

In star formation studies, the refinement 
criterion when using an AMR mesh is usually based on the Jeans length. In other words, the Jeans length needs to be adequately sampled to properly resolve 
the system dynamics.
There has been some debate as to how many cells per Jeans length are actually necessary, and authors commonly use 10-16 cells per Jeans length 
\citep[e.g.][]{Commercon2011a,Krumholz2012}. \citet{Vaytet2016} recently showed, using 1D 
simulations, that resolution can affect the thermodynamics of collapsing dense clouds, because of poor sampling of the optical depth which limits radiation 
cooling and causes spurious heating inside the first Larson core. If the optical depth within a cell is too large (typically $>100$), \citet{Vaytet2016} found that the radiative flux points inward the first core, which creates a spurious bump in the temperature profile. This numerical effect happens when the numerical resolution is too low, and  \citet{Vaytet2016} showed empirically that limiting the optical depth within a cell to a few tens is enough to prevent it.
 We performed a resolution study to show that this effect can be also prevented in 3D simulations and to ensure it was not affecting the evolution of the protostellar system.

\begin{figure}
\centering
\includegraphics[width=0.48\textwidth]{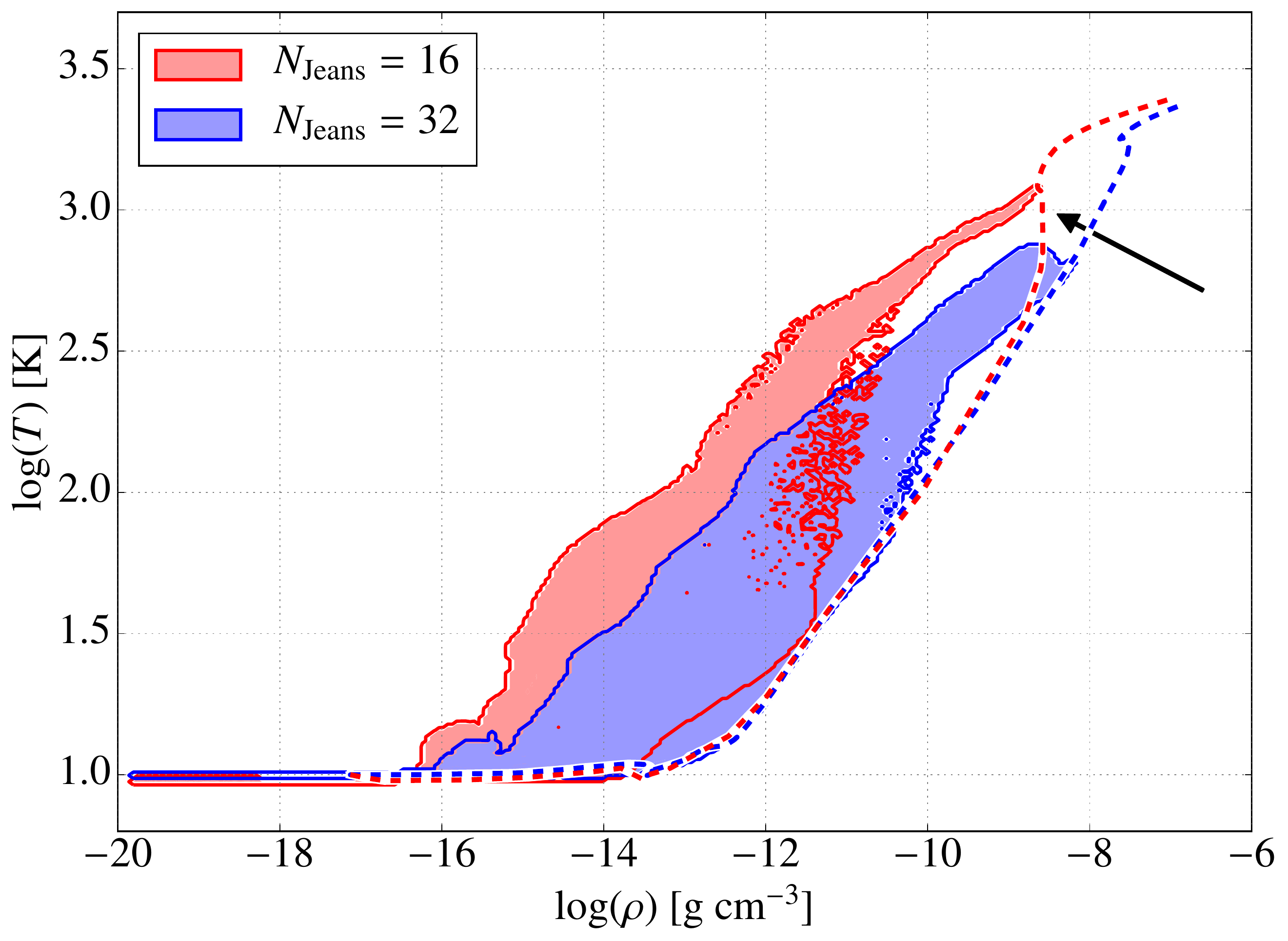}
\caption{Temperature as a function of density, for every cell in the computational domain (ideal MHD case). The 
simulation using $N_{\text{Jeans}} = 32$ cells per Jeans length is represented by the blue area, while the red region is for the run with only 16 cells 
per Jeans length. Each data set is delineated by a solid contour line which outlines the data distributions. The two snapshots were taken at similar evolution 
times, chosen to be just after the $N_{\text{Jeans}} = 16$ run has departed from its initial adiabatic track. The dashed lines represent the time evolution of the 
central (densest) cell inside the mesh (these tracks continue beyond the time of the snapshots to provide a wider context). The black arrow indicates the place 
where the low-resolution track departs from its original adiabat.}
\label{fig:resolution}
\end{figure}

To determine the resolution requirements of our set-up, we ran a simulation with a lower resolution of 16 cells per Jeans length and compare it to our fiducial 
resolution of 32 cells per Jeans length. The results are shown in Fig~\ref{fig:resolution}. The red contours are for the low-resolution run, while the blue 
contours are for the calculation with 32 cells 
per Jeans length. The dashed lines show the evolution of the densest cell in the system, and can be compared to the 1D results of \citet{Vaytet2016}. In the 
low-resolution run, we actually observe a `turn off' in the first adiabatic phase, at densities $\sim10^{-9}~\gcc$, while the high-resolution path continues 
along the same adiabatic track. This departure from adiabaticity actually looks identical to the phenomenon observed by \citet{Vaytet2016}. We also note that 
the gas is hotter in the low-resolution simulation. It is obvious here that 16 cells per Jeans length is not enough to properly describe 
the physical processes at work. In fact, we can also see just at the top right end of the high-resolution track a small `kink' in the curve, suggesting that 
even 32 cells might not be enough for fully converged results. However, the simulation with ambipolar and ohmic diffusion would have been too expensive to 
run with anything more than $N_{\text{Jeans}} = 32$, and we determined that the consequences of such a small kink would only be minimal.

From this short resolution study, we see that a refinement criterion solely based on the local Jeans length is not adapted to describe the adiabatic evolution of 
a hydrostatic core in collapse calculations. A dedicated study of the necessary numerical resolution within the different components of a collapsing core 
(envelope, disk, hydrostatic cores) is clearly needed and should be the focus of future work.

\section{Regions of active ambipolar and ohmic diffusion}\label{app:elsasser}

\begin{figure}
\centering
\includegraphics[width=0.48\textwidth]{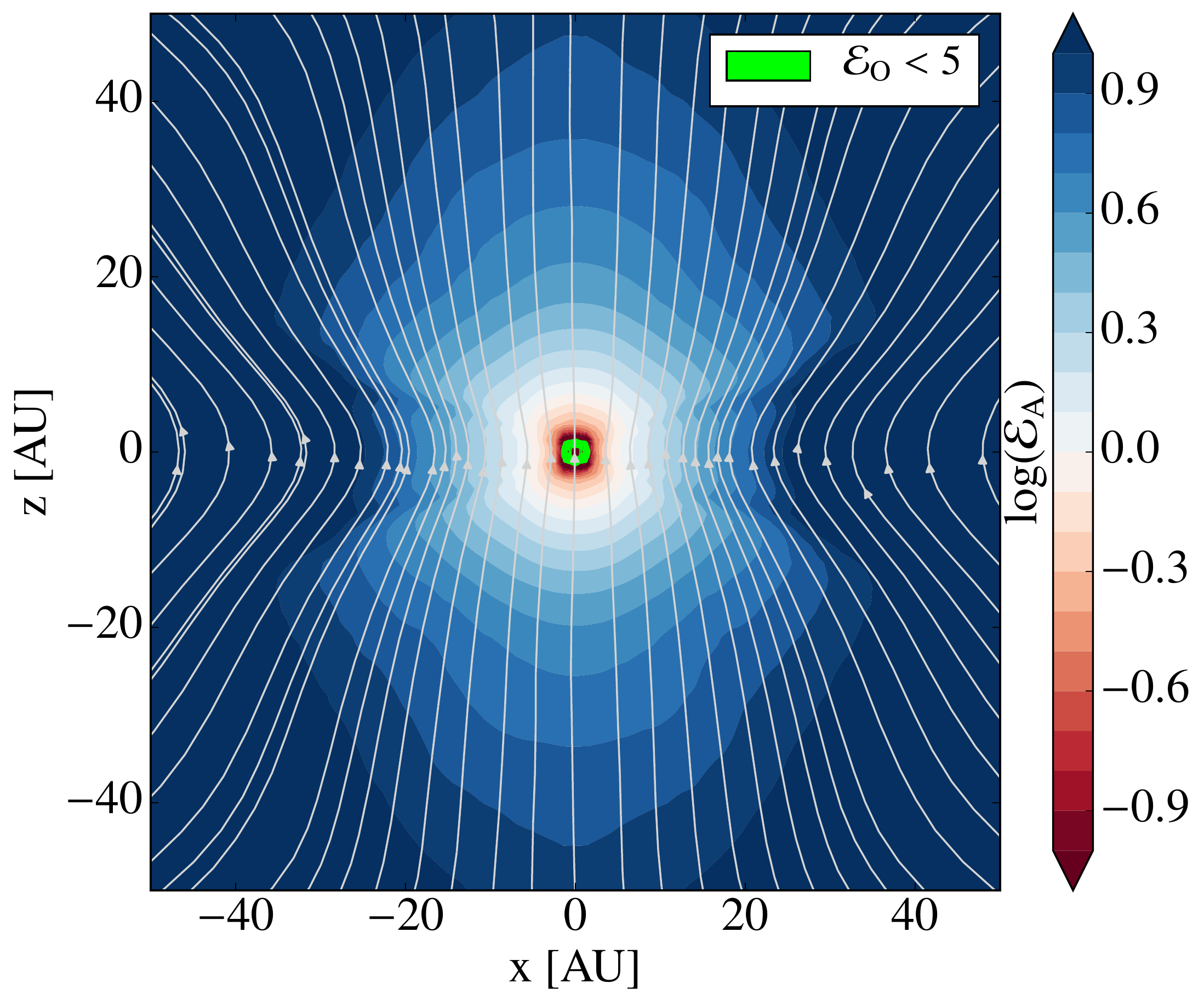}
\caption{Map of the ambipolar (filled blue/red contours) and ohmic (green) Reynolds numbers close to the first Larson core. The light grey lines represent the 
magnetic field.}
\label{fig:elsasser}
\end{figure}

We compute here dimensionless numbers which reveal the regions on active ambipolar and ohmic diffusion in our system.
Following \citet{Tomida2015,Masson2016}, we define the ambipolar and ohmic Reynolds (or sometimes called Elsasser) numbers as
\begin{equation}
\mathcal{E}_{\text{A}} =  \frac{VL}{\eta_{\text{A}}} ~~;~~ \mathcal{E}_{\text{O}} =  \frac{VL}{\eta_{\text{O}}} ~,
\end{equation}
where $V$ is the magnitude of the gas velocity vector, and $L$ represents the typical scale of the system, which we take as the distance from the current cell
to the centre of the protostar. Figure~\ref{fig:elsasser} shows a map of the logarithm of $\mathcal{E}_{\text{A}}$ (coloured contours) in the vicinity of the first 
Larson core (side view) with the magnetic field lines overlayed (light grey). The regions where $\mathcal{E}_{\text{A}} \lesssim 1$ (white and red) have strong ambipolar 
diffusive effects that modify the magnetic field topology. Indeed, the equatorial pinching of field lines, which is evident in the IMHD run
(see Fig.~\ref{fig:slices}m), is reduced when $\mathcal{E}_{\text{A}} < 5$ (inside 30 AU), and eventually disappears when $\mathcal{E}_{\text{A}} < 1$ (inside 10 AU).

In contrast, the green region in Fig.~\ref{fig:elsasser} represents areas where ohmic diffusion is active ($\mathcal{E}_{\text{O}} < 5$); it is much smaller because 
the ohmic resistivities peak at higher densities than their ambipolar counterpart (see Fig.~\ref{fig:brho_ministep}d). This reveals that the straightening of the 
field lines observed in Sec.~\ref{sec:morph_1stcore} and Fig.~\ref{fig:slices}f,m is due to the effects of ambipolar diffusion.

\section{Definitions of the first and second proto-stellar cores}\label{app:cores}

\begin{figure}
\centering
\includegraphics[width=0.48\textwidth]{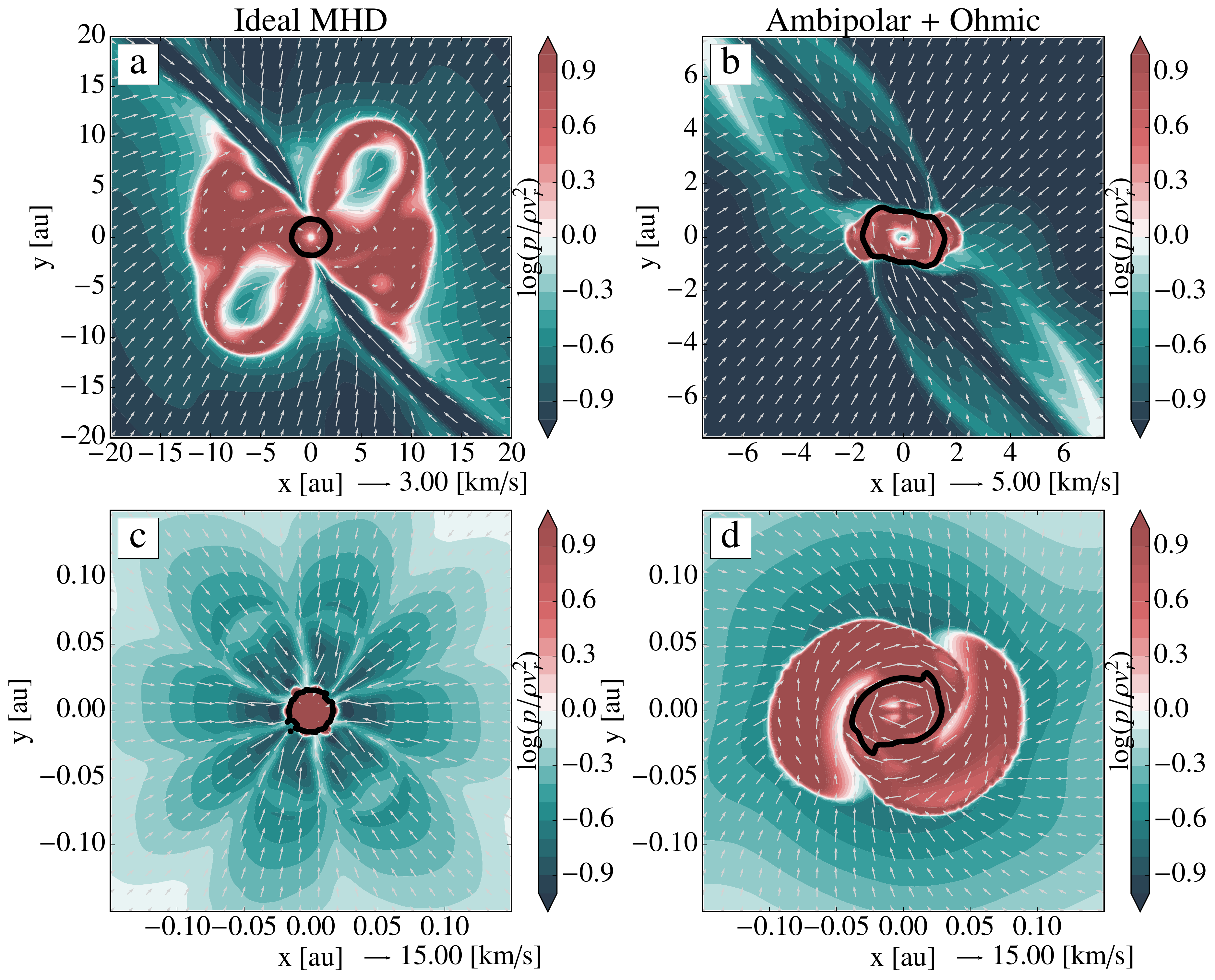}
\caption{Maps and contours showing the morphologies of the cores using two different definitions. The coloured maps show the ratio of thermal to infalling ram 
(kinetic) pressure, while the black solid contour defines the region where the gas density exceeds density thresholds of $\rho_{\text{core}} = 10^{-10}~\gcc$ for 
the first Larson core and $\rho_{\text{core}} = 10^{-5}~\gcc$ for the second Larson core. The panels are: (a) \texttt{runID} first core, (b) \texttt{runAO} first 
core, (c) \texttt{runID} second core, (d) \texttt{runID} second core. We note the difference in spatial scales between panels (a) and (b).}
\label{fig:cores}
\end{figure}

In this section, we take a look at two different definitions of the first and second Larson cores and how they may affect core morphologies, masses and radii.
The cores are often referred to as `hydrostatic cores' in the literature, as they are supposedly (for the most part) in hydrostatic equilibrium. Computing the
condition for hydrostatic equilibrium is often expensive in a 3D system, as pressure gradients have to be calculated in all directions, and authors have often 
favoured simpler criteria such as vanishing radial velocities or thermal-to-kinetic pressure equilibrium. Choosing one definition over the other can sometimes
result in large differences in the extent of the core, and consequently the mass that is attributed to it. In Fig.~\ref{fig:cores}, we compare two different 
definitions for the proto-stellar cores. These are:
\begin{enumerate}
\item Thermal pressure exceeds ram pressure: $p > \rho v_{\text{r}}^{2}$
\item Density exceeds a chosen threshold: $\rho > \rho_{\text{core}}$
\end{enumerate}
The first condition characterises a thermally supported body, and is equivalent (within a factor of $\gamma$) to the definition in \citet{Tomida2010}.
The second definition is the one we have used throughout this paper. We chose $\rho_{\text{core}} = 10^{-10}~\gcc$ for the first Larson core and
$\rho_{\text{core}} = 10^{-5}~\gcc$ for the second Larson core.

The left column of Fig.~\ref{fig:cores} shows the first and second cores in \texttt{runID}, while the right one is for \texttt{runAO}. For the first core in
\texttt{runID} (panel a), it is clear that definitions 1 is affected by the interchange instability which creates a large region of thermally supported 
gas. The resulting morphology is not what is usually associated with a hydrostatic core, with loops presumably connected to the magnetic field. On the other hand, 
definition 2 yields a close-to-spherical body. In contrast, both definitions produce similar results for the \texttt{runAO} first core (panel b), where the core 
is an unbroken/continuous body, flattened on its north and south faces by the heavy accretion streams that slam onto its surface. In the case of the
second core, the situation is reversed. Both definitions agree for \texttt{runID} (panel c) but large discrepancies emerge for \texttt{runAO} (panel d).
Indeed, the small disk around the second core is also pressure-supported (see Sec.~\ref{sec:late_evol}) and definition 1 considers it to be part of the 
proto-stellar core, while definition 2 selects only a small spheroidal core, excluding the disk around it.

\end{document}